\documentclass{aastex63}
\usepackage{enumitem,kantlipsum}

\submitjournal{The Astrophysical Journal}

\shorttitle{Configuration of Sparking Discharges in Pulsars}

\begin{document}

\title{Two Dimensional Configuration and Temporal Evolution of Sparking discharges in Pulsars}


\author[0000-0003-1824-4487]{Rahul Basu}
\affiliation{Janusz Gil Institute of Astronomy, University of Zielona G\'ora, ul. Szafrana 2, 65-516 Zielona G\'ora, Poland.}

\author[0000-0003-1879-1659]{George I. Melikidze}
\affiliation{Janusz Gil Institute of Astronomy, University of Zielona G\'ora, ul. Szafrana 2, 65-516 Zielona G\'ora, Poland.}
\affiliation{Evgeni Kharadze Georgian National Astrophysical Observatory, 0301 Abastumani, Georgia.}

\author[0000-0002-9142-9835]{Dipanjan Mitra}
\affiliation{National Centre for Radio Astrophysics, Tata Institute of Fundamental Research, Pune 411007, India.}
\affiliation{Janusz Gil Institute of Astronomy, University of Zielona G\'ora, ul. Szafrana 2, 65-516 Zielona G\'ora, Poland.}

\begin{abstract}
We have investigated the evolution of a system of sparking discharges in the 
inner acceleration region (IAR) above the pulsar polar cap. The surface of the
polar cap is heated to temperatures around $10^6$ K and forms a partially 
screened gap (PSG) due to thermionic emission of positively charged ions from 
the stellar surface. The sparks lag behind the co-rotation speed during their 
lifetimes due to variable \textit{\textbf{E}}\textbf{x}\textit{\textbf{B}} 
drift. In a PSG the sparking discharges arise in locations where the surface 
temperatures go below the critical level ($T_i$) for ions to freely flow from 
the surface. The sparking commences due to the large potential drop developing 
along the magnetic field lines in these lower temperature regions and 
subsequently the back streaming particles heat the surface to $T_i$. The 
temperature regulation requires the polar cap to be tightly filled with sparks 
and a continuous presence of sparks is required around its boundary since no 
heating is possible from the closed field line region. We have estimated the 
time evolution of the sparking system in the IAR which shows a gradual shift in
the spark formation along two distinct directions resembling clockwise and 
anti-clockwise motion in two halves of the polar cap. Due to the differential 
shift of the sparking pattern in the two halves, a central spark develops 
representing the core emission. The temporal evolution of the sparking process 
was simulated for different orientations of the non-dipolar polar cap and 
reproduced the diverse observational features associated with subpulse
drifting.
\end{abstract}

\keywords{pulsars:}

\section{Introduction}
\noindent
The pulsar magnetosphere is characterised by a steady outflow of relativistic 
plasma along the open magnetic field lines. The majority of the electromagnetic
radiation from pulsars arises within this outflowing plasma, including the
radio emission that results from non-linear plasma instabilities \citep{MGP00,
GLM04,LMM18,RMM20}. The outflowing plasma is also the source of the dense 
pulsar wind responsible for the pulsar wind nebulae \citep{dJ07,KCL15}. 
Detailed estimates have constrained the the radio emission to originate from
heights of less than 10 percent of the light cylinder radius \citep{vHX97,KG98,
ET_MR02,ML04,WJ08,KMG09,M17}. Thus the location of the plasma generation region
is well inside the inner magnetosphere. The prototype for the plasma generation
region is the inner vacuum gap (IVG) above the polar caps, extending to heights
of $\sim$ 10-100 meters from the stellar surface, and potential difference of 
10$^{12}$ V across it \citep{S71,RS75}. The outflowing plasma is generated as a
series of sparking discharges due to creation of $e^-e^+$ pairs in presence of 
high magnetic fields and are accelerated to relativistic speeds in opposite 
directions by the large potential drop, setting up a cascading effect. The 
charges within each spark screen the electric potential till they leave the gap
due to inertial motion, when the subsequent sparks are formed resulting in a 
non-stationary plasma flow. 

There are certain limitations in the vacuum gap model of the inner 
acceleration region (IAR) above the polar cap. The back-flowing relativistic
charges in the sparks heat up the surface to temperatures $\sim$ 10$^{6}$ K 
which is close to the critical level ($T_i$) for ionic\footnote{We are 
considering stars where $\bf\Omega$\textbf{$\cdot$}\textit{\textbf{B}} $<0$ and
have positive charge densities above the polar caps such that a IAR gap can be 
established.} free flow from the surface \citep{CR80,J86}. As a result the IAR 
is not a complete vacuum but forms a partially screened gap (PSG) with a steady
density of positively charge ions. It was shown by \citet{GMG03} that in a PSG,
where the ionic density ($\rho_i$) are as high as 90 percent of the 
Goldrich-Julian co-rotational density \citep[$\rho_{GJ}$,][]{GJ69}, sparking 
discharges can develop in the IAR. The spark formation is governed by the 
surface temperature ($T_S$) with thermostatic self-regulation of the potential 
drop in the IAR. In a purely vacuum gap there is presence of unscreened 
potentials between sparks and hence the discharges cannot be confined at any 
location on the surface, but is expected to scatter in the direction opposite 
to the principal normal of the curvature of the local magnetic field lines 
\citep{CR80}. The thermostatic self-regulation in a PSG ensures that the 
sparking is limited to finite size with a typical length scale. 

Other constraints on the physical properties of the IAR has been obtained from 
several detailed studies. The surface magnetic field above the polar cap is 
expected to be highly non-dipolar in nature. The efficiency of the pair cascade
leading to the sparking discharge as well as the high multiplicity ($10^5$) of 
the outflowing plasma requires magnetic field lines with radius of curvature 
$\sim10^5-10^6$ cm \citep{TH19}, in contrast to $\sim10^8$ cm for purely 
dipolar fields. The non-dipolar nature of the polar cap has also been supported
by measurements of the thermal X-ray emission from the hot polar cap surface as
well as simultaneous radio/X-ray studies of pulsars \citep{GHM08,H13,SGZ17,G17,
HKB18,AM19,SG20,PM20}. The sparks are expected to undergo variable 
\textit{\textbf{E}}\textbf{x}\textit{\textbf{B}} drift in the IVG which results
in the well known observational behaviour of subpulse drifting \citep{WES06,
BMM16} in the pulsed radio emission. It has been suggested that the sparks in 
the IAR rotate around the magnetic axis \citep{RS75,SvL17}, however, in our 
previous work \citep[][hereafter Paper I]{BMM20b} it was shown that in the 
absence of any external electric field, the sparks in IAR would lag behind 
co-rotation speed during their lifetimes. The phase behaviour associated with 
drifting periodicity reflects the dynamics of the sparking process in the IAR 
\citep{BM18a,BMM19}. In Paper I the lagging behind motion of the sparks along 
with the non-dipolar nature of the surface fields was used to simulate 
different categories of phase behaviour associated with drifting \citep{BMM19}. 

One important observational feature of the drifting behaviour is the absence of
drifting in the central core region of the profile window \citep{R86,BMM19}, 
and cannot be explained by lagging behind motion of the sparks throughout the 
polar cap. It was recognized in Paper I that the lagging behind scenario is 
restricted by the presence of polar cap boundaries beyond which no spark can be
formed in the closed field line region. The thermal regulation of the hot polar
cap surface requires a continual presence of sparking discharges along the 
boundary since no additional heating is possible from the closed field lines on
the other side. In this work we have expanded on the ideas presented in Paper I
by including the effects of the polar cap boundary on the sparking 
configuration. A two-dimensional model of the sparking discharges in a PSG is 
presented and we explore its evolution with time. The constraints from the 
polar cap boundary coupled with the lagging behind co-rotation motion of the 
sparks during their lifetimes introduce two distinct trajectories for the 
temporal evolution of the sparking distribution. One half exhibits a clockwise 
evolution while the other half shows an anti-clockwise behaviour. The 
differential shift of the sparking pattern in the two halves as well as the 
tightly packed nature of sparking distribution implies that heating location at
the center is stationary in most configurations of non-dipolar polar caps and 
accounts for the absence of drifting in the core component.

In the following sections we present the details of the model and simulations 
of tightly packed sparking distribution in the IAR starting with the physical 
characteristics of sparks in section \ref{sec:sparkphysics}. Section 
\ref{sec:sparkbound} demonstrates the effect of the polar cap boundary on the 
location of the subsequent sparks and their shifts to be in opposite directions
in two halves of the polar cap. The two-dimensional configuration of the 
sparking discharges and their temporal evolution is explored in section 
\ref{sec:spark2D}. We have used the estimated sparking distribution and its 
time evolution to simulate the subpulse drifting behaviour. The drifting 
behaviour for different orientations of the surface non-dipolar magnetic fields
is shown in section \ref{sec:sparkdrift}. A short discussion summarising the 
implications of this work is presented in section \ref{sec:conclusion}.

\section{Physical parameters of Sparks in a Partially Screened Gap}\label{sec:sparkphysics}
\noindent
The details of the spark formation in a PSG was first presented in 
\citet{GMG03} and have been further explored in several subsequent works 
\citep[][Paper I]{GMZ06,S13,SMG15,MBM20}. In this section we summarize the 
different physical parameters like the spark size, the time scales associated 
with the sparking process, etc., that has been estimated in these earlier works
and will be used later to understand the sparking configuration in the IAR.

\subsection{Size of spark}
\noindent
The charge density above the polar cap is limited by the co-rotational density, 
$\rho_{GJ}=\Omega B/2\pi c$, which is many orders of magnitude lower than the 
density of ions on the stellar crust. As a result a sufficiently heated polar 
cap can sustain a free flow of ions from the tail end of the surface charge 
distribution at the critical Temperature, $T_i \approx \epsilon_i/30k$, here 
$\epsilon_i$ is the binding energy of ions and $k$ is the Boltzmann constant 
\citep{CR80}. For surface temperatures, $T_S < T_i$, a charge depleted 
acceleration region is formed which is populated by ions with density $\rho_i <
\rho_{GJ}$. A sparking discharge is setup in this region which heats the 
surface till $T_S = T_i$ is reached and the ionic free flow is restored 
terminating the spark. The effective potential drop $\Delta V_{PSG}$ across 
each spark is given as :
\begin{equation}
\Delta V_{PSG} = \frac{4 \pi \eta b B_d \cos{\alpha_l}}{P c} h_{\perp}^2
\end{equation}
here $\eta = 1 - \rho_i/\rho_{GJ} = 1 - \exp[30(1 - T_i/T_S)]$, $b = B_s/B_d$ 
where $B_s$ is the non-dipolar surface magnetic field and $B_d = 2\times10^{12}
(P \dot{P}_{-15})^{0.5}$ G, the equivalent dipolar field, $\alpha_l$ is angle
between the local magnetic field and the rotation axis, $h_{\perp}$ is the 
perpendicular radius of the spark and $P$ and $\dot{P}$ are the pulsar rotation 
period and period derivative respectively. The total energy deposited per unit 
area on the surface of the polar cap by the back streaming electrons can be
estimated by multiplying the energy of each electron with the particle flux
in the spark $P_S = (e \Delta V_{PSG})\times(\eta n_{GJ} c)$. Equating the 
deposited energy per unit area with the energy radiated from the heated 
surface, $P_S = \sigma T^4$, the typical radius of a spark can be estimated as
\begin{equation}
h_{\perp} = 260 ~~\frac{T_6^2}{\eta b (\cos{\alpha_l})^{0.5}} \left(\frac{P}{\dot{P}_{-15}}\right)^{0.5} {\rm cm,} 
\label{eq:hperp}
\end{equation}
where the surface temperature $T_S = T_6\times10^6$ K.

\subsection{Timescales associated with Sparking}
\noindent
There are two relevant time scales associated with the sparking process, the
heating time ($t_s$) or spark duration, the time taken by the sparking process to heat the surface to the critical temperature, and the cooling time ($t_c$) 
which is the time elapsed between the cessation of the sparking process and the
surface to cool sufficiently for the next spark to commence. These estimates 
require understanding the physical properties of the neutron star crust where 
the heating is taking place. At temperatures around million Kelvin the crust 
permeated by highly non-dipolar magnetic fields have been shown to exist in a 
crystalline state and the surface density ($\rho_s$) can be calculated in such 
cases as \citep{L01} :
\begin{equation}
\rho_s \simeq 561 A Z^{-3/5} b^{6/5}~~{\rm g~cm^{-3}},
\end{equation}
Where $A$ is the mass number and $Z$ the atomic number of the constituent atoms
in the crust. The crust is made up of iron atoms with $A = 56$ and $Z = 26$ and
the density is $\rho_s = 4.45\times10^3 b^{6/5}$ g~cm$^{-3}$. The durations of 
heating and cooling of the crust depends on the depth of heat deposition and 
can be estimated using the radiation length $y = 14$ g~cm$^{-2}$ for iron ions.
The corresponding depth is given as
\begin{equation}
L_R = y/\rho_s \simeq 3.15\times10^{-3}~b^{-6/5}~~{\rm cm}.
\end{equation}
The heat can be deposited up to a depth of several times $L_R, L \approx 
10^{-3}$ cm \citep{GMG03}. 

The total energy carried by the back-streaming particles is used to heat the 
crust with specific heat per unit volume $C_H$ to a depth of $L$, $C_H L 
\partial T/\partial t \simeq C_H L T/t_s$, assuming uniform heating. The 
specific heat of the crust has contribution from the lattice vibrations as well
as the free electrons and is given as :
\begin{equation}
C_H \approx 4.4\times10^{12} \rho_6 (1 + 0.024\rho_6^{-2/3} T_6) \approx 2\times10^{10}~b^{6/5} (1 + 0.89 b^{-4/5}T_6)~~{\rm erg~K^{-1}~cm^{-3}}.
\end{equation}
The sparking timescales can be estimated from the above expressions as :
\begin{equation}
t_s \approx 30~(1 + 0.89~b^{-4/5}~T_6)~T_6^{-3}~~\mu{\rm sec}.
\label{eq:sprktime}
\end{equation}
The sparks lag behind co-rotation velocity and the total distance during the 
lifetime is $h_s = \eta v_{cr} t_s$. The co-rotation velocity was estimated in
Paper I to be around 10$^6$ cm~s$^{-1}$ resulting in $h_s \approx$ 10~cm.

Finally, in order to estimate the cooling time scale the heat transport 
equation is used
\begin{equation}
C_H \frac{\partial T}{\partial t} = \frac{\partial}{\partial l}\left(\kappa\frac{\partial T}{\partial l}\right)
\end{equation}
where $\kappa$ is the thermal conductivity of the crust. The surface layer for
heat penetration is very thin and can be considered to have a uniform heat 
conductivity. The partial differential equation can be approximated as 
$\partial T/\partial t \approx T/t_c$ and $\partial^2 T/\partial^2 l \approx 
T/L^2$. This gives an order of magnitude estimate of the cooling time scale as
\begin{equation}
t_c = \left(\frac{L^2 C_H}{\kappa}\right)
\label{eq:sprkcool}
\end{equation}
In the outermost layers of the neutron star the heat transport is primarily
dominated by the electronic transfer and for $b\sim10$ the estimated 
$\kappa\sim10^{12}$ erg~cm$^{-1}$~s$^{-1}$~K$^{-1}$ \citep{GMG03}. Hence, the 
cooling timescale of the polar cap is around 100 $n$sec which is more than two 
orders of magnitude shorter than the duration of sparks. It is possible for the 
subsequent sparking process to commence near the previous spark soon after it 
dies as the surface cools almost instantaneously.

\section{Spark formation and effect of Polar Cap boundary} \label{sec:sparkbound}

\subsection{Sparking in IAR}
In the PSG model the polar cap is positively charged 
($\bf\Omega$\textbf{$\cdot$}\textit{\textbf{B}} $<0$) with the surface 
temperatures maintained around the critical level for free flow of ions ($T_i$)
due to thermostatic regulation from sparking discharges. When the surface 
temperature ($T_S\sim10^6$ K) is above $T_i$ there is a steady outflow of 
positively charged ions from the stellar surface with density $\rho_{GJ}$, 
screening the potential drop along the IAR. If the temperature becomes lower 
than $T_i$ the ionic density drops below $\rho_{GJ}$ and a large potential drop 
appears along the IAR. As a result sparking discharges with cascading $e^--e^+$
pairs are setup, with the pairs separated by the large potential difference. 
The electrons are accelerated downwards and heat the surface while the 
positrons are accelerated away from the surface and give rise to the outflowing
plasma. The sparks grow in size till sufficient pairs are produced and the 
surface is heated to $T_S \gtrsim T_i$, thereby screening the potential drop 
across the IAR and terminating the spark. The typical lifetime of a spark is 
around 30 $\mu{\rm sec}$ (Eq.\ref{eq:sprktime}) during which the spark grows 
both vertically and laterally ($h_{\perp}$ in Eq.\ref{eq:hperp}) forming a 
cylindrical plasma column. The point of maximum heating lies at the center of 
the spark and falls off gradually towards the edge with no sharp boundaries in 
the temperature distribution below the spark. It has been shown in Paper I that
the sparking plasma column during their lifetimes lag behind the co-rotation 
motion of the pulsar. As a result the maximally heated point at the termination
of the spark is shifted by a small distance $h_s$ opposite to the co-rotation 
direction. The dense plasma column after the termination of sparking leaves the
IAR in about $\sim$ 300 $n$sec which is comparable to the cooling time of the 
surface (Eq.\ref{eq:sprkcool}) and is much shorter than the duration of the 
sparks. In the absence of any other constraints the subsequent spark forms 
immediately at the location of maximum heating (hence fastest cooling), shifted
by a distance $h_s$, resulting in an apparent drift motion.

If the size of the polar cap is larger than the lateral size of a spark then 
thermal regulation of the surface requires the presence of a system of 
tightly packed sparks in the IAR. There are no hard borders expected between 
adjacent sparks and the region near their edges are heated by particles 
from all surrounding sparks. The polar cap has a well defined boundary 
separating the open and closed magnetic field line regions. The closed field 
line region has constant $\rho_{GJ}$ charge density and hence it is not 
possible to have sparking in the closed field line region with the boundary 
cutting across a spark, i.e. no spark can be formed straddling the boundary 
between the open and closed field line regions. The continuous heating of the 
surface around the boundary region requires an annular band of tightly packed 
sparks to be formed closely bordering the boundary. In order to ensure the
effective heating of the surface to continue, no gaps can appear in the normal 
direction away from the boundary to the nearest spark during the lagging behind
evolution of the sparking process. As a result the lagging behind motion of the
sparking pattern is constrained by the polar cap boundary and subsequent sparks
can only evolve around the annular ring either in the clockwise or 
anticlockwise manner (see section \ref{subsec:lagpol}). If there is space for 
more than one ring of sparks in the polar cap, the presence of the outer band 
constrains the inner sparks to also be arranged in a nested inner concentric 
ring touching the outer sparking band in a closely packed setup. The effective 
heating of the surface requires the sparks in the inner ring to not have any 
gaps with the outer ring in the normal direction, thereby constraining the 
subsequent sparking pattern to evolve around the inner ring. Depending on the 
availability of space, multiple contiguous rings of increasingly smaller radius
will be formed in this setup with sparks constrained to evolve around each 
ring, till there is space for only a single spark at the center. There is no 
space for evolution of sparking at the center and after the previous spark 
heats the surface above $T_i$ and is extinguished, the next spark is formed 
around the same location after the surface cools down. The size of the central 
spark depends on the available space at the center and can be smaller than a 
typical spark. The arrangement of sparks in concentric rings around a central 
spark is consistent with the core-cone nature of the average radio emission 
beam \citep{ET_R90,ET_R93,GKS93}, where the pulsar emission is arranged in 
concentric rings of conal emission around a central stationary core. The 
lifetimes of individual sparks are much shorter than the apparent drift motion 
of the sparking pattern which evolve over several seconds. Hence, the observed 
drifting behaviour reflects the change in the sparking pattern and is not 
associated with motion of individual sparks.

\subsection{Effect of Polar Cap boundary on Sparking}\label{subsec:lagpol}

\begin{figure}
\gridline{\fig{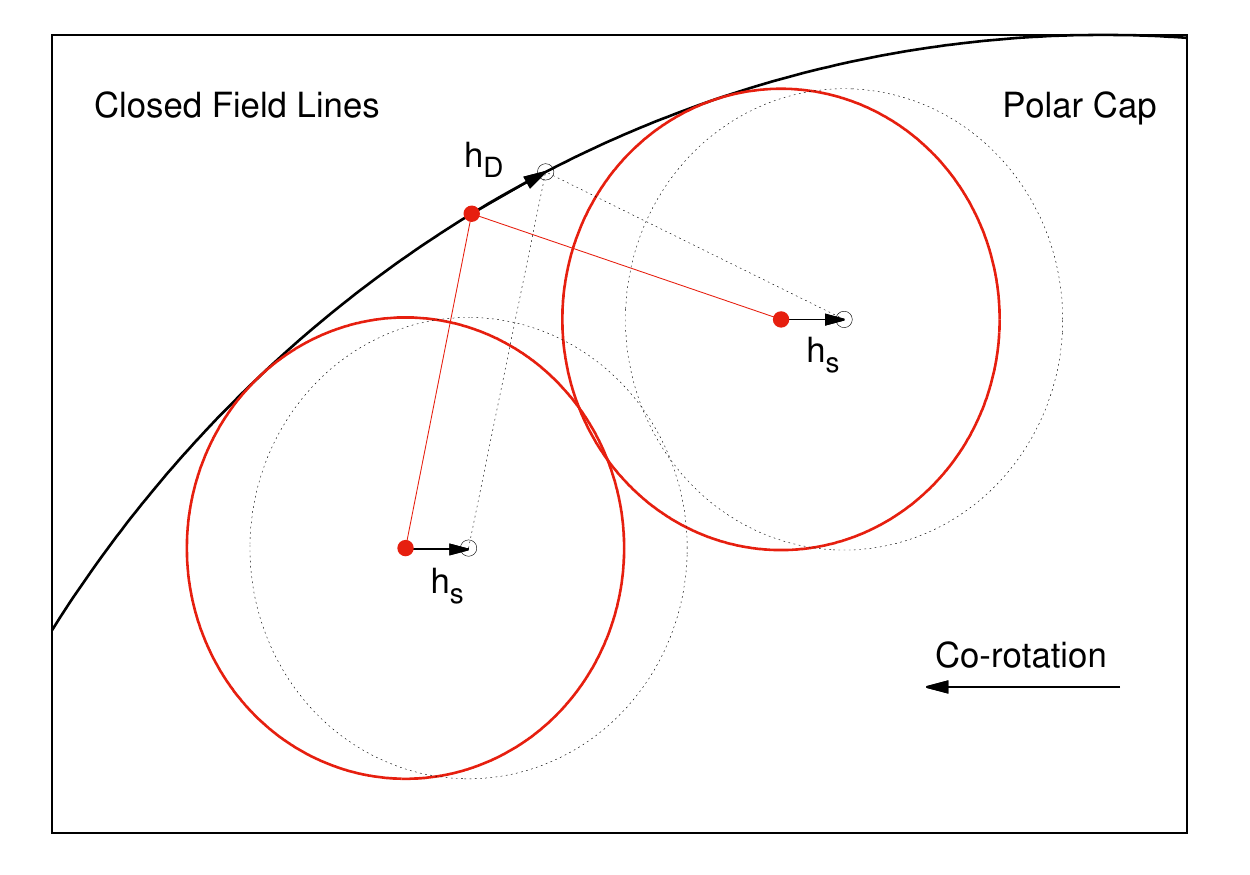}{0.6\textwidth}{(a)}}
\gridline{\fig{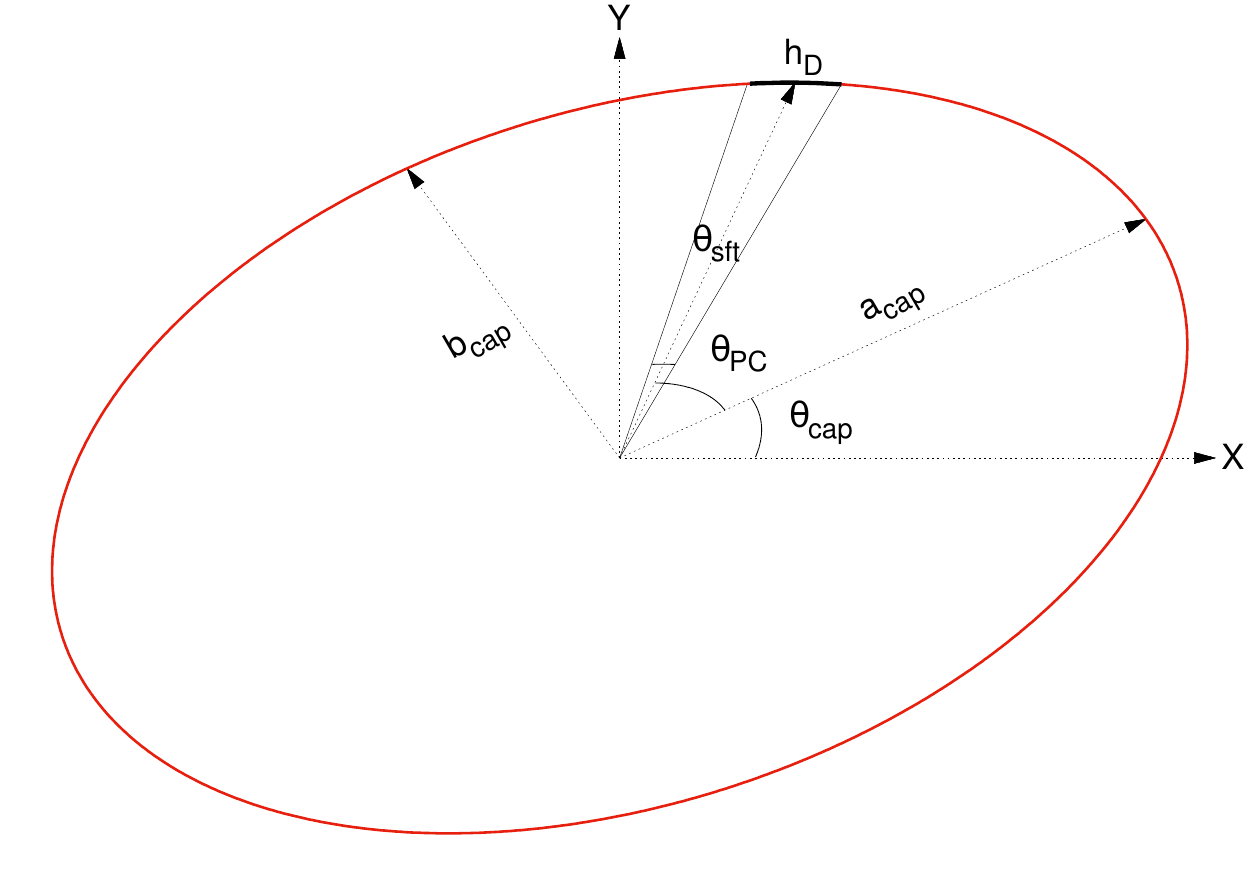}{0.55\textwidth}{(b)}}
\caption{(a) The upper panel shows a schematic of the evolution of the sparking
pattern along the polar cap boundary due to drifting of the sparks. The black 
curved line is the boundary of the polar cap where two sparks (red circles) 
develop around a maximally heated point (red dot) on the boundary. The sparks 
drifts by a distance $h_s$ during their lifetimes as they lag behind the 
co-rotation direction. The maximally heated point (open black circle) at the 
end of the sparking process shifts by a distance $h_D$ due to the drifting of 
the sparks. After the surface cools down the subsequent sparks are formed 
around the shifted point (open black circle) resulting in a shift of the 
sparking pattern. (b) The lower panel shows a schematic of the elliptical polar
cap with major axis $a_{cap}$, minor axis $b_{cap}$ and inclined at an angle 
$\theta_{cap}$ in the x-y plane. The spatial shift $h_D$ along the boundary of 
the polar cap at a point specified by the polar angle $\theta_{PC}$ corresponds
to the angular shift $\theta_{sft}$.
\label{fig:sparkshift}}
\end{figure}

Figure \ref{fig:sparkshift}a shows a schematic of the evolution of the sparking 
process along the boundary. Initially the sparks (red outline) are formed close
to each other in a tightly packed configuration around a point (red point) on 
the boundary which is maximally heated by particles from both sparks. During 
their lifetimes the sparks shift by an average distance $h_s$ opposite to the 
co-rotation direction (dotted black outline). At the end of the life cycle the 
location of the maximally heated point (open black circle) shifts around the 
boundary due to the drift motion. After the surface cools down the next sparks 
are formed around this shifted point in the boundary resulting in a effective 
shift of the sparking pattern by a distance $h_D$ along the boundary of the 
polar cap.

\begin{figure}
\gridline{\fig{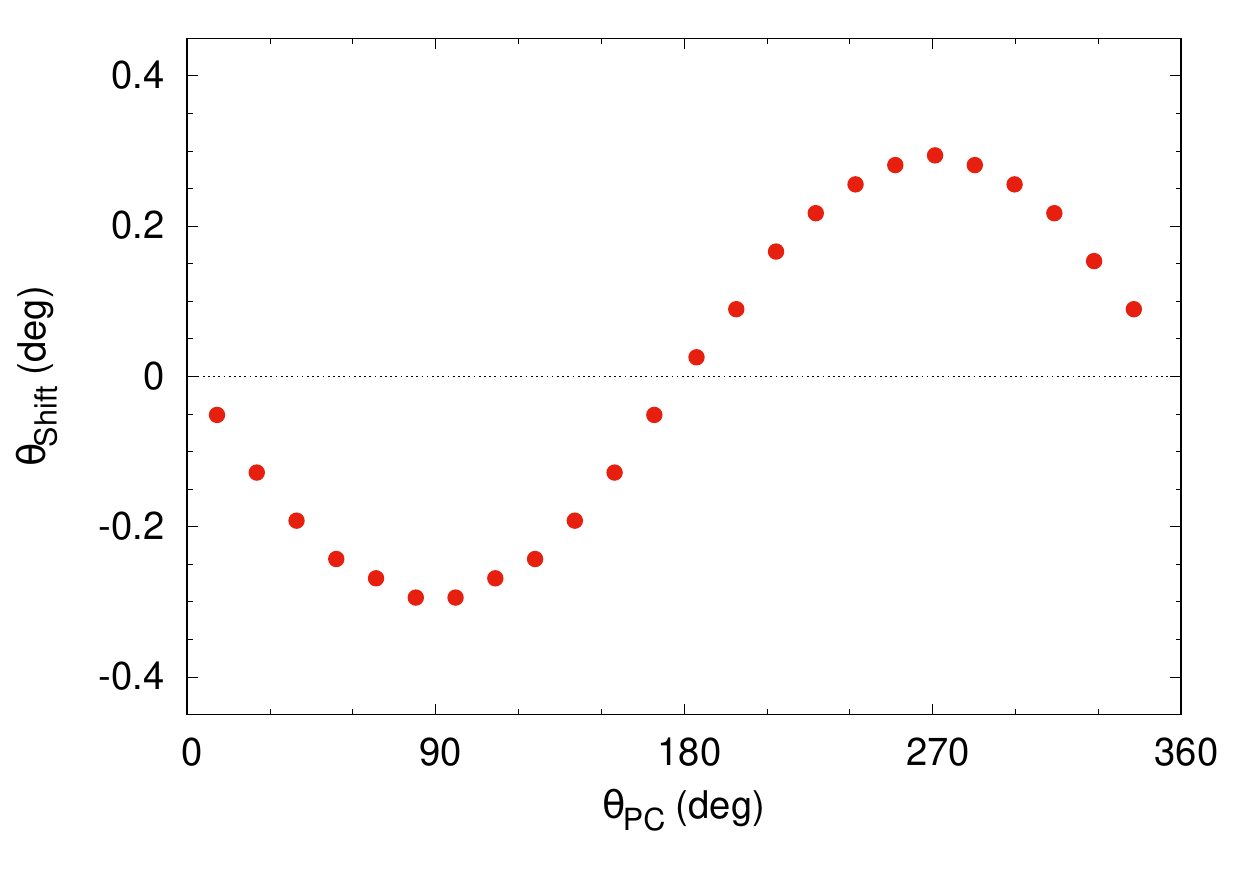}{0.5\textwidth}{(a)}
          \fig{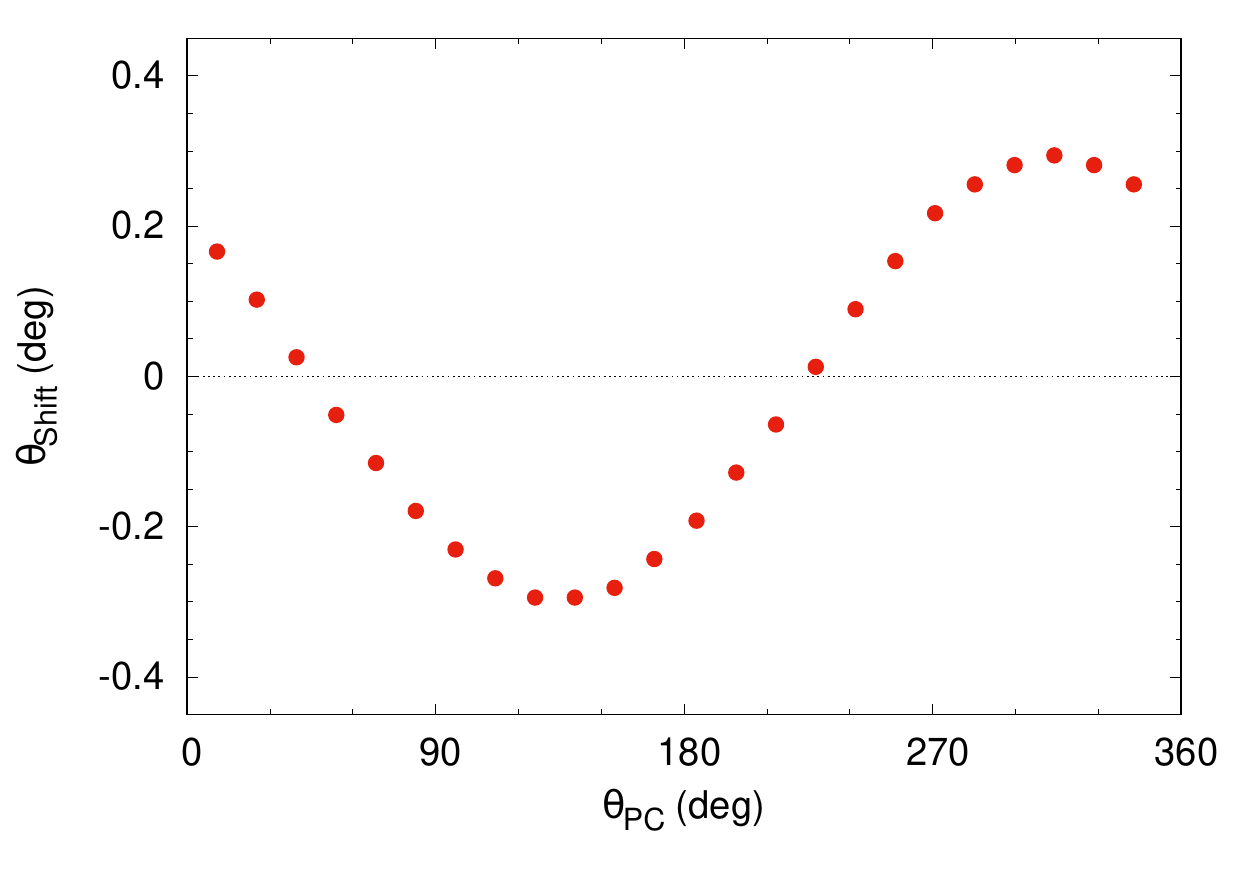}{0.5\textwidth}{(b)}}
\gridline{\fig{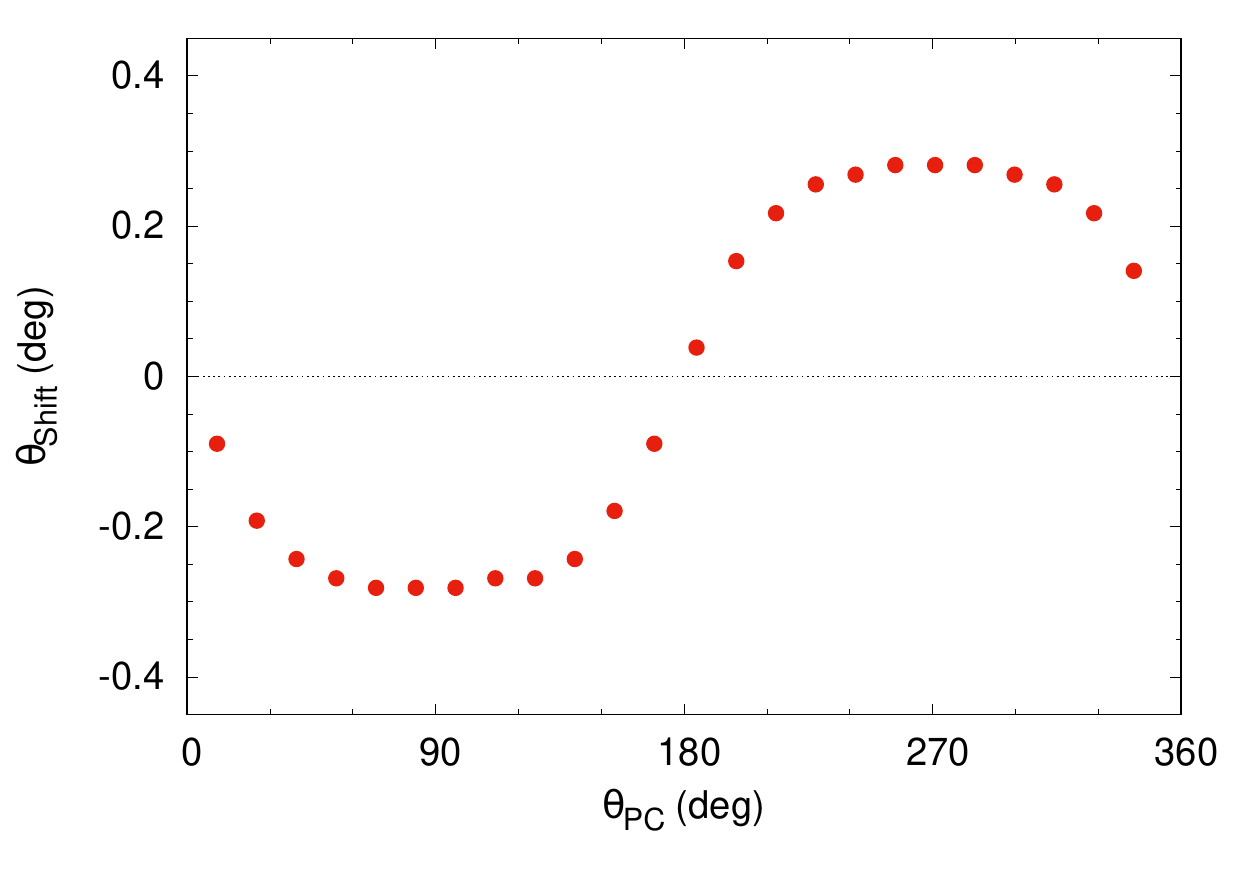}{0.5\textwidth}{(c)}
          \fig{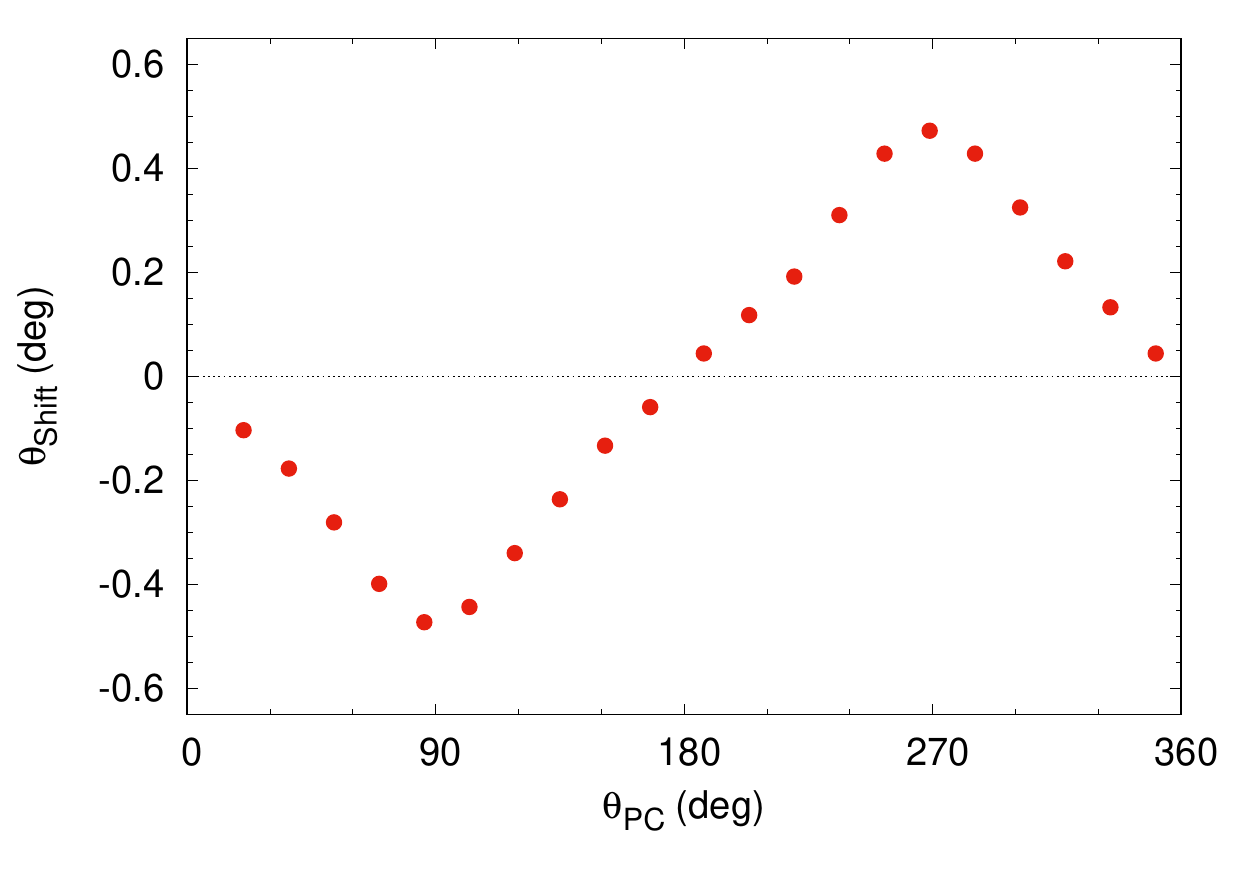}{0.5\textwidth}{(d)}}
\caption{The figure shows the angular shift in the sparking pattern around the
entire boundary of the polar cap ($\theta_{PC}$). The panels in (a) and (b) 
represent circular polar caps where the sparks lag behind the co-rotation 
direction along the x-axis in case (a), while the sparks lag behind co-rotation
along an axis which makes a 45$\degr$ angle in the x-y plane in case of (b). In
panels (c) and (d) examples of elliptical polar caps are shown with 
eccentricity 0.6 and sparks lagging behind along the x-axis in both cases. In 
panel (c) the major axis is oriented along the x-axis while in panel (d) the 
major axis is oriented along the y-axis.
\label{fig:surfshift}}
\end{figure}

The angular shift ($\theta_{sft}$) in the drift direction of the sparking 
pattern at different points on the boundary is related to the spatial shift 
($h_D$) at a point specified by the polar angle ($\theta_{PC}$) as :
\begin{equation}
\theta_{sft} = h_D/\sqrt{(a_{cap}^2\sin^2{\theta_{PC}} + b_{cap}^2\cos^2{\theta_{PC}}}, 
\end{equation} 
here $a_{cap}$ and $b_{cap}$ are the major and minor axes of the elliptical 
polar cap with $a_{cap} = b_{cap} = R_{cap}$ for a circular case. The shifts in
the sparking pattern along the entire polar cap boundary is shown in Figure 
\ref{fig:surfshift} corresponding to, (a) circular polar cap with the sparks 
lagging behind co-rotation along the x-axis, (b) circular polar cap with the 
co-rotation direction making a 45$\degr$ angle in the x-y plane, (c) elliptical
polar cap with eccentricity 0.6 and major axis along the x-axis, and (d) 
elliptical polar cap with eccentricity 0.6 and major axis along the y-axis. The
shift is maximum when the direction of co-rotation is tangential to the 
curvature of the boundary and goes to zero when co-rotation is normal to the 
boundary. In all cases there are two distinct regions of sparking evolution 
bounded by the points where the the co-rotation direction is normal to the 
elliptical/circular boundary of the polar cap. In one half the shifts are in 
the negative direction which signify a clockwise shift of the sparking pattern, 
while in the other half the shifts are positive resembling an anti-clockwise 
shift. In the elliptical polar caps the maximum shifts are flatter or steeper 
compared to the circular case depending on the orientation of the major axis. 
In realistic situations the differential shifts will be averaged out due to 
finite size of the sparks, as the next spark can only form in the available 
space between two sparks and an average shift will be seen over time. We expect
two distinct evolution pattern of the sparking discharges along the rim of the 
polar cap showing clockwise behaviour in one half and anti-clockwise evolution 
in the other half. We have used this insight about the evolution pattern to 
understand the two dimensional configuration of the sparking system.

\section{The Sparking Configuration and its Temporal Evolution} \label{sec:spark2D}
\noindent
The distribution of sparks on the polar cap surface in the presence of a PSG 
have the following constraints: 
\begin{enumerate}[leftmargin=*]
\item A continuous presence of sparking is required bordering the boundary of 
the polar cap for effective thermal regulation.
\item The sparks are as tightly packed as possible within the IAR.
\item There are two distinct direction of sparking evolution with one half 
showing a clockwise shift in the pattern and the other half an anti-clockwise 
shift.
\end{enumerate}
The first condition requires the presence of an annulus of sparks bordering the
boundary of the polar cap in a tightly packed configuration. The second 
condition ensures that in the interior of the polar cap the sparks are also
formed in concentric layers with a single spark at the center. The central 
spark may differ in size depending on the available space. Finally, the third 
condition associated with subpulse drifting requires the sparking pattern in 
each concentric layer to shift in two different directions in two halves 
resembling a clockwise and anti-clockwise shift. The bounding points of the two
layers are locations where the lagging behind co-rotation direction is normal 
to the concentric ring, the starting point being the inward normal and the 
ending point the outward normal. As the sparking pattern shifts away/towards 
the bounding points smaller empty spaces may open up in these regions. A spark 
appears in these locations due to the temperature going below the critical 
level, and as a consequence the potential drop appears, to start the sparking 
process. But there is heating in the surrounding regions from the sparks on 
either side of this location and hence the potential drop across it is lower, 
resulting in smaller sparks. Due to differential motion around it the central
spark is not expected to show any shift but reappear around the same location 
depending on the surface heating requirements.

In a purely dipolar magnetic field configuration the polar cap boundary can be 
approximated to be circular in shape. However, as discussed in the introduction
a number of detailed observations have revealed the polar cap to be non-dipolar
in nature. In this work the non-dipolar nature of the surface field is modelled
by considering a combination of magnetic dipoles of different strengths (see 
section \ref{sec:sparkdrift}). A large star centered dipole is used to 
reproduce the largescale magnetic field away from the stellar surface, while a 
weaker dipole is placed close to the polar cap to match the surface field 
strength as well as the estimated polar cap size \citep{GMM02,BMM20b}. In this 
configuration the boundary of the non-dipolar polar cap is elliptical in shape 
with the direction of elongation depending on the location of the surface 
dipole. Although the magnetic field configuration considered in this work 
provides a convenient setup for estimating the non-dipolar surface magnetic 
field, it is not an unique solution. But the sparking evolution presented here 
should be applicable to polar caps with well defined continuous boundaries. An 
elliptical polar cap is defined by a major axis ($a_{cap}$), a minor axis 
($b_{cap}$) and angle of inclination of the major axis in the x-y plane 
($\theta_{cap}$). The potential drop across the sparks are expected to vary 
along the different axes in elliptical polar caps and we approximate the spark 
shape to resemble the polar cap with major axis $a_{sprk}$, minor axis 
$b_{sprk} = a_{sprk}b_{cap}/a_{cap}$ and inclination angle $\theta_{cap}$, such
that $\sqrt{a_{sprk} b_{sprk}} \approx h_{\perp}$. The sparks are unlikely to 
have sharp boundaries but are expected to have a peak density and spread out 
till the surface is heated by an adjacent spark. The number of tracks of the 
concentric sparking trajectories in addition to the central core is $N_{trk} = 
{\rm Int}({a_{cap}/a_{sprk}})$. The maximum number of sparks that can be 
accommodated within any concentric annulus within the polar cap is given as :
\begin{eqnarray}
N_{sprk}^i = {\rm Int}\left(F (a_{out}^i b_{out}^i - a_{in}^i b_{in}^i)/(a_{sprk} b_{sprk}))\right). \nonumber\\
a_{out}^i = a_{cap} - 2(i-1) a_{sprk},~a_{in}^i = a_{out}^i - 2a_{sprk}, \nonumber\\
b_{out}^i = b_{cap} - 2(i-1) b_{sprk},~b_{in}^i = b_{out}^i - 2b_{sprk}. 
\end{eqnarray}
Here $i = 1,2...N_{trk}$ and $F$ is a scaling factor for maximum packing which 
we find to be around 0.75. A series of sparks are setup along each of these 
concentric regions whose angular size is $\theta_{sprk}^i = 2\pi/N_{sprk}^i$, 
and their centers lie on the ellipse specified by $a_{trk}^i = (a_{out}^i + 
a_{in}^i)/2$ and $b_{trk}^i = (b_{out}^i + b_{in}^i)/2$. As specified before
the dynamics of the evolution of the sparking pattern follows two different 
directions in two halves. In one half the next sparks are formed shifted by an 
angle $\theta_{u}^i = -h_{D}/\sqrt{a_{trk}^i b_{trk}^i}$ and in the other half 
by the angle $\theta_{d}^i = +h_{D}/\sqrt{a_{trk}^i b_{trk}^i}$, here $h_{D}$ 
being the average shift of the sparking pattern which is constant for all $i$. 
Finally, smaller sparks are set up at either ends whose size is variable 
depending on the available space between the two trajectories.

\begin{figure}
\begin{interactive}{animation}{circ_sprkcap.mp4}
\gridline{\fig{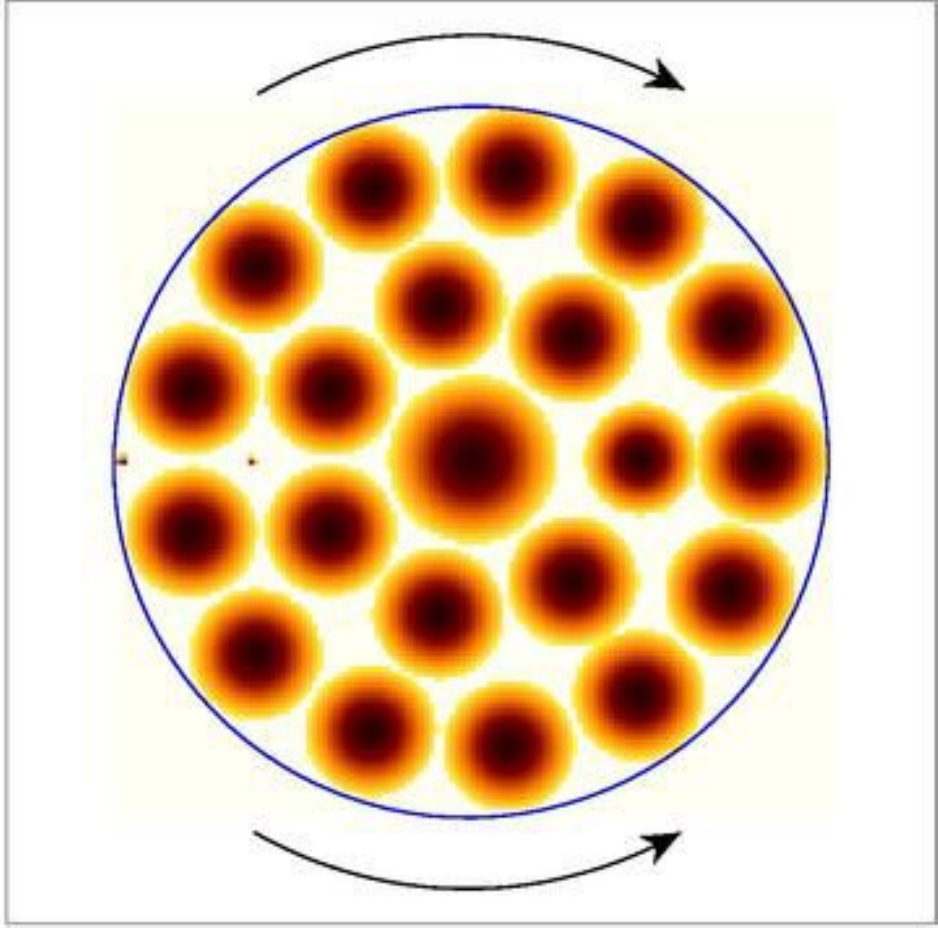}{0.48\textwidth}{(a)}
          \fig{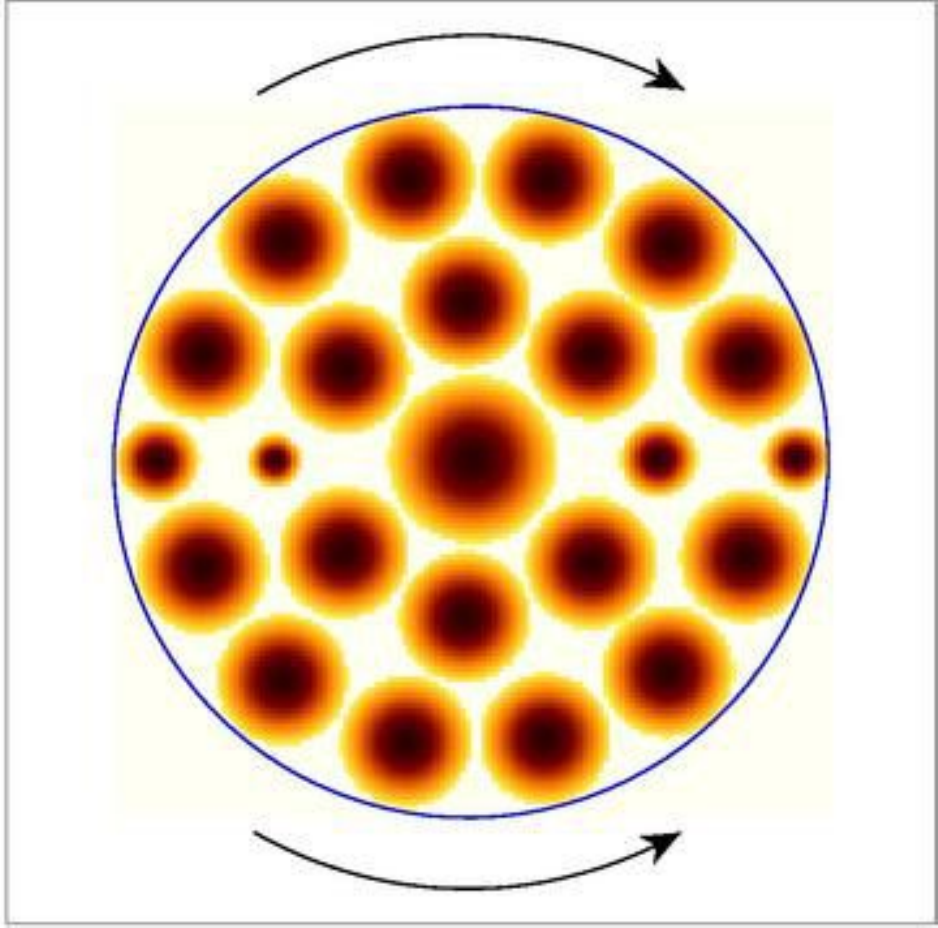}{0.48\textwidth}{(b)}}
\end{interactive}
\caption{The figure shows two different snapshots of the distribution of 
sparking on a circular polar cap. The sparks are arranged within two concentric
rings from the boundary and contains a central spark to form a tightly packed 
configuration. The sparking pattern evolves with time to show a clockwise shift
in the upper half and a anti-clockwise shift in the lower half (see arrows) 
around the central spark. Smaller sparks are formed at the edges when the 
available space is less than the spark size but the surface still requires 
thermal regulation. The sparks in the figures do not have sharp borders, which 
are drawn primarily as guidelines to show the distinction between adjacent 
sparks. \\\\ An animation showing the evolution of the sparking configuration 
with time is available.
\label{fig:spark2d}}
\end{figure}

The two dimensional sparking configuration for a circular polar cap, where 
$a_{cap} = b_{cap}$, with lagging behind co-rotation direction along the x-axis
is shown in Figure \ref{fig:spark2d}. The sparking pattern evolves to resemble 
a clockwise shift in the upper half and an anti-clockwise shift at the lower 
half. The two regions are separated at the points where the co-rotation 
direction is normal to the circle, i.e, at $\theta_{PC} = 0\degr, 180\degr$ 
where $\theta_{PC}$ is represents the polar angle. As the spark pattern from 
either side shift away from $\theta_{PC} = 180\degr$, a gap opens up which in 
certain cases is smaller than the size of a fully formed spark and smaller size
sparks are formed. Similarly, at $\theta_{PC} = 0\degr$, the available space 
reduces due to the encroachment from either side leading to smaller sparks. We 
have assumed that no phase difference exist between the shifts in the upper and
lower halves. However, the sparks in these two regions evolve independently and
can have any arbitrary phase difference between them.

\section{Modelling Subpulse Drifting} \label{sec:sparkdrift}
\noindent
The evolution of the sparking distribution in the IAR is seen as the phenomenon
of subpulse drifting in the pulsar radio emission. We have used the model of 
the two dimensional sparking distribution presented in the previous section to 
simulate the different drifting behaviour observed in pulsars. The process of
generating a single pulse sequence from a pulsar with non-dipolar polar cap has
been presented in Paper I and is summarized below. 

The non-dipolar magnetic field in the polar cap is modelled using a large star 
centered dipole which dominate the magnetic field away from the star surface 
along with one or more surface dipoles which determine the magnetic field 
structure at the polar cap \citep{GMM02}. We use a spherical co-ordinate system
with the origin located at the center of the neutron star and the rotation axis
along $\theta=0\degr$. The star centered dipole located at the origin is 
specified as \textit{\textbf{d}} = ($d, \theta_d, 0\degr$), where $d$  
represents the dipole moment and $\theta_d$ the inclination angle between the
rotation and magnetic axis of the star. The surface dipoles are represented as
\textit{\textbf{m$_i$}} = ($m^i, \theta_m^i, \phi_m^i$), where i = 1, 2, ..., N
in case more than one is used, with dipole moments $m^i = 0.001-0.05d$ much 
smaller than the star centered dipole, and the orientation of each dipole 
specified by the angles $\theta_m^i$ and $\phi_m^i$. The surface dipoles are 
located at \textit{\textbf{r$_i$}} = ($r_s^i, \theta_s^i, \phi_s^i$) close to 
the surface with $r_s^i = 0.95 R_S, R_S = 10$~km is the neutron star radius, 
while the angles $\theta_s^i$ and $\phi_s^i$ can be varied to shift the 
location away from the purely dipolar polar cap. At a distance of $\sim30R_S$ 
the contribution of surface dipole moments becomes negligible and the magnetic 
field is largely dipolar with contribution from \textit{\textbf{d}}. The 
opening angle corresponding to the last open dipolar field line at $30R_S$ is 
used to estimate the outline of the modelled polar cap surface by numerically 
solving the magnetic field line equations (see Paper I for details). The 
elliptical outline of the polar cap is estimated using non-linear Least Square 
fits \citep{PTV92} to obtain $a_{cap}$, $b_{cap}$ and $\theta_{cap}$ and the 
location of the center defined by $\theta_{cap}^c$, $\phi_{cap}^c$. 

As the pulsar rotates the pulsed emission is seen when the line of sight (LOS) 
passes through the open field line across the radio emission region. The pulsar
profile shape and the drifting behaviour is dependent on the LOS cut across 
this emission beam. The LOS is defined by the angle $\beta$ which is the 
minimum angle between the axis of the star centered dipole and LOS. In the 
spherical co-ordinate system defined at the center of the neutron star, with 
the rotation axis aligned along the z-axis, the track of the LOS and 
consequently a train of pulsed emission is obtained by continuous change in the
coordinate $\phi=2\pi t/P$, for a fixed azimuth angle $\theta = 
\theta_d+\beta$. The non-dipolar polar cap is usually shifted from the dipolar 
case with the magnetic field lines twisting as they connect with the dipolar 
emission region around $10-100R_S$. As a result the outline of the LOS gets 
modified as it traverses the sparking distribution resulting in different 
drifting behaviour for different surface magnetic field configurations. Once 
the LOS enters the open field line region the corresponding field line at the 
surface is estimated by numerically solving the magnetic field line equations. 
The sparking intensity at that point on the surface is estimated and recorded. 
It is assumed that the radio emission intensity follows that of the sparking, 
but in realistic cases the radio emission in the subpulses are a result of 
non-linear plasma processes \citep{MGP00} which introduces additional features.
As the LOS traverses the extent of the pulse window an intensity pattern due to
the sparking distribution is obtained and forms the single pulse. In subsequent
rotations the sparking distribution evolves and gives different intensity 
patterns, reflecting the drifting behaviour. 

In order to estimate the two dimensional sparking distribution a Cartesian 
$x'y'$-plane\footnote{The primed co-ordinates are used to distinguish from the 
star centered co-ordinate system used for estimating the LOS traverse.} is 
defined to contain the elliptical polar cap with origin at the center of the 
ellipse. The boundary of the upper and lower halves of the polar cap signifying
clockwise and anti-clockwise evolution of the sparking pattern is specified by 
the angles $\theta'_s = 3\pi/2 - \phi_{cap}^c$ and $\theta'_e = \pi/2 - 
\phi_{cap}^c$, here $\theta'$ is the polar angle in the $x'y'$-plane. The 
sparking pattern diverges away from $\theta'_s$ and converges toward 
$\theta'_e$. We have assumed a Gaussian distribution of intensity for each 
spark with elliptical symmetry :
\begin{equation}
I_{sprk} (x', y') = I_0\exp{\left(-\frac{1}{2}\left[(x' - x'_c)^2/a_{sprk}^2 + (y' - y'_c)^2/b_{sprk}^2\right]\right)} 
\end{equation}
Here $I_0$ is the peak intensity of spark and $x'_c$, $y'_c$ corresponds to the
center of the spark. The sparks do not have sharp borders and the intensity 
pattern is drawn primarily as guidelines to show the distinction between 
adjacent sparks. The location of the sparks within the elliptical polar cap and
the evolution of their distribution with time is described in the previous 
section.

Similar to Paper I we have considered three different magnetic field 
configurations whose behaviour represent the three distinct observational 
classes of subpulse drifting, viz., coherent phase-modulated drifting, phase 
stationary drifting and Bi-drifting with opposite drift directions in different
parts of the emission window \citep[see][]{BMM19}. We present below the 
physical characteristics of the PSG for each magnetic field configuration, the 
temporal evolution of the two dimensional sparking configuration and the 
simulated single pulses showing subpulse drifting.

\subsection{Coherent Phase-Modulated drifting}
\noindent
Coherent phase-modulated drifting is the prototype of the subpulse drifting 
behaviour with the subpulses showing a systematic shift throughout the entire 
pulse window. These are mostly associated with LOS traverses towards edge of 
the emission beam and seen in pulsar average profiles with single or double 
conal components \citep{R86,BMM19}. In Paper I the magnetic field configuration
used to demonstrate this behaviour consisted of a star centered dipole 
\textit{\textbf{d}} = ($d, 15\degr, 0\degr$) and one surface dipole 
\textit{\textbf{m}} = ($0.001d, 0\degr, 0\degr$) located at \textit{\textbf{r}}
= ($0.95R_S, 18.86\degr, 10.99\degr$). The dipolar magnetic axis is tilted by 
an angle 15$\degr$ with the rotation axis while the non-dipolar polar cap is 
located around 5$\degr$ away in the $\phi$-axis from the dipolar polar cap. 

\begin{deluxetable}{ccccccccc}
\tablecaption{The physical parameters of elliptical polar cap \label{tab:capfit}}
\tablewidth{0pt}
\tablehead{
    & \colhead{$a'_{cap}$} & \colhead{$b'_{cap}$} & \colhead{$\theta'_{cap}$} & \colhead{$\theta_{cap}^c$} & \colhead{$\phi_{cap}^c$} & \colhead{$b$} & \colhead{$\cos{\alpha_l}$} &\colhead{$\eta$} \\
   & \colhead{(m)}  & \colhead{(m)} & \colhead{($\degr$)} &  \colhead{($\degr$)} & \colhead{($\degr$)} &   &   &   }
\startdata 
  Coherent & 127.6 & 80.4 & 45.9 & 16.2 & 5.3 & $\sim4$ & $\sim0.85$ & 0.038 \\
   &   &   &   &   &   &   &   &   \\
 Phase-Stat & 75.3 & 21.6 & 52.4 & 51.2 & 17.9 & $\sim19$ & $\sim0.8$ & 0.021 \\
   &   &   &   &   &   &   &   &   \\
  Bi-drift & 62.2 & 36.1 & 148.5 & 3.2 & 100.1 & $\sim26$ & $\sim0.79$ & 0.013 \\
\enddata
\end{deluxetable}

The polar cap is elliptical in shape with the fitting parameters shown in Table
\ref{tab:capfit}. The surface magnetic field strength is characterized by 
average $b\sim4$ (see Fig. 17~in Paper I). We are primarily interested in the 
polar caps of long period older pulsars which show subpulse drifting. A number 
of studies have shown that the average emission beam in such cases comprise of 
a central core surrounded by two concentric rings of conal emission 
\citep{ET_R90,ET_R93,MD99,ET_MR02}. The presence of two conal rings in such 
pulsars is also expected from the spark sizes obtained from the PSG model 
\citep{MBM20}. Hence, using $N_{trk} =2$, the typical spark size can be 
estimated as $a_{sprk}\sim23.6$ m, $b_{sprk}\sim14.9$ m and $h_{\perp}\sim18.8$
m. The screening factor for the PSG is obtained from eq.(\ref{eq:hperp}) as 
$\eta\sim0.038$. The drifting periodicity ($P_3$) in a PSG is given by $P_3 = 
1/2\pi\eta\cos{\alpha_l}$, where $\alpha_l$ is the angle of the local 
non-dipolar magnetic field with the rotation axis \citep{MBM20}. The average 
$\cos{\alpha_l}$ is 0.85 for the magnetic field configuration (Fig. 17~in Paper
I) and we can estimate $P_3$ to be 4.9$P$. 

\begin{figure}
\begin{interactive}{animation}{coherent_sprkcap.mp4}
\epsscale{0.55}
\plotone{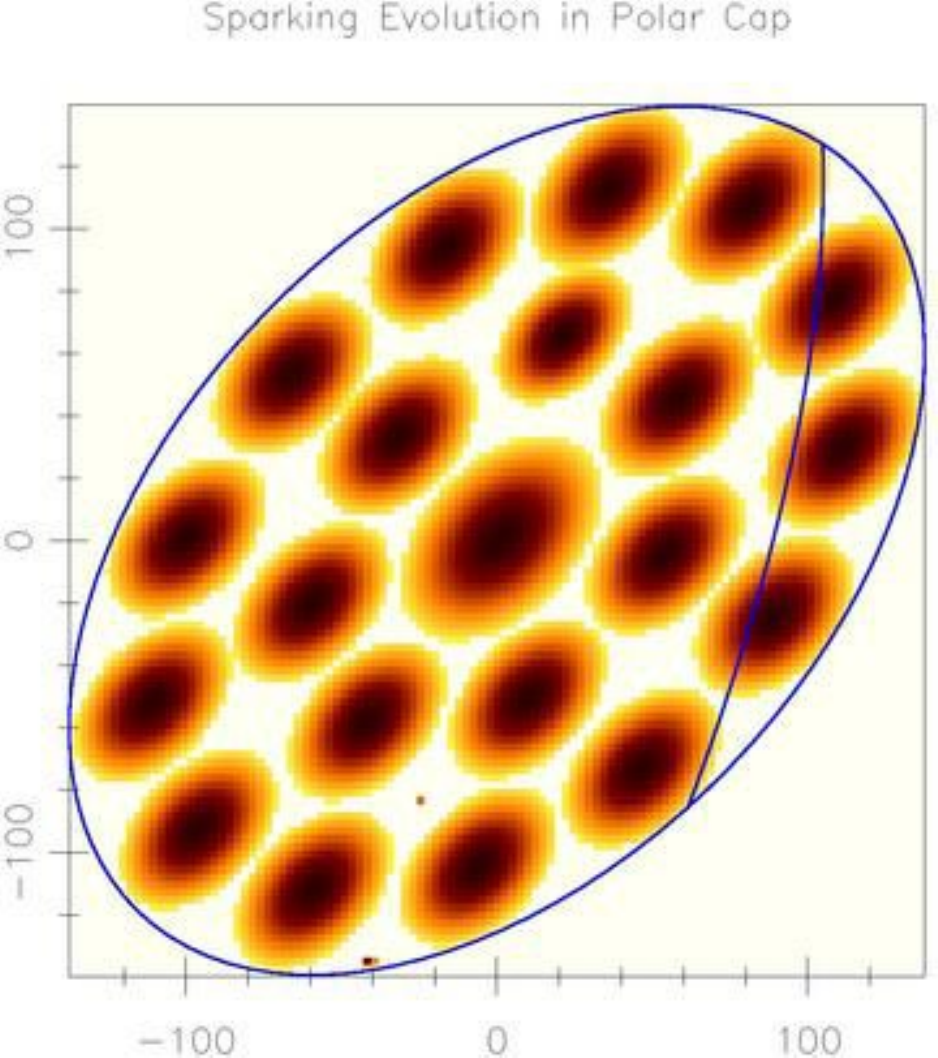}
\end{interactive}
\caption{The figure shows the two dimensional distribution of the sparking 
pattern for an elliptical polar cap with major axis of 127.6 m, minor axis of 
80.35 m and tilted by an angle of 45.9$\degr$ in the plane of the ellipse. The 
sparks are arranged in two concentric elliptical annulus around a central spark
in a tightly packed configuration. The sparking pattern evolves with time to 
show a clockwise shift in the left half and an anti-clockwise shift in the 
right half bounded by the points $\theta'_s = 264.7\degr$ and $\theta'_e = 
84.7\degr$, where the pattern shifts away from $\theta'_s$ and converges 
towards $\theta'_e$. The line of sight (LOS) traverse across the emission beam 
at an angle $\beta=3\degr$ from the center and its imprint on the polar cap is
also shown. The dynamical evolution of the sparking distribution across the LOS
results in coherent subpulse drifting. \\\\ An animation showing the evolution 
of the sparking configuration with time is available.
\label{fig:sparkcoh}}
\end{figure}

\begin{deluxetable}{ccccccccccc}
\tablecaption{The details of the sparking distribution in polar cap exhibiting coherent drifting \label{tab:sparkcoh}}
\tablewidth{0pt}
\tablehead{
    & \colhead{$i$} & \colhead{$a_{out}$} & \colhead{$a_{in}$} & \colhead{$b_{out}$} & \colhead{$b_{in}$} & \colhead{$N_{sprk}$} & \colhead{$\theta_{sprk}$} & \colhead{$a_{trk}$} & \colhead{$b_{trk}$} & \colhead{$\omega_{u,d}$}  \\
   &   & \colhead{(m)} & \colhead{(m)} & \colhead{(m)} & \colhead{(m)} &   & \colhead{($\degr$)} & \colhead{(m)} & \colhead{(m)} & \colhead{(deg s$^{-1}$)} 
          }
\startdata
  Outer & 1 & 127.6 & 80.4 & 80.4 & 50.6 & 13 & 27.7 & 104.0 & 65.5 & $\mp$5.6 \\
  Inner & 2 & 80.4 & 33.2 & 50.6 & 20.8 & 7 & 51.4 & 56.8 & 35.7 & $\mp$10.43 \\
\enddata
\end{deluxetable}

The two dimensional distribution of the sparks and their evolution with time is
shown in Fig.\ref{fig:sparkcoh}. The sparks are distributed around two 
elliptical rings around a central spark in a tightly packed configuration.
The details of the sparking distribution for each track is reported in Table 
\ref{tab:sparkcoh} and shows the different major ($a$) and minor ($b$) axes 
describing each ring, the maximum number of fully formed spark that can be 
accommodated along these rings ($N_{sprk}$), the angular size of the sparks 
($\theta_{sprk}$), as explained in section \ref{sec:spark2D}. The two bounding
points around which the sparking distribution show evolution in opposite 
directions are $\theta'_s = 264.7\degr$ and $\theta'_e = 84.7\degr$, where the 
pattern shifts away from $\theta'_s$ and converges towards $\theta'_e$. The 
rate of shifting of the patterns in the two halves of each ring is estimated as
$\omega_{u,d}$ = $\mp\theta_{sprk}/P_3$. The average shift of the sparking 
pattern during a rotation period ($P=1$ s) is $h_D=|\omega_{u,d}|\sqrt{a_{trk}
b_{trk}}\sim8.1$ m (see section \ref{sec:spark2D}).

\begin{figure}
\gridline{\fig{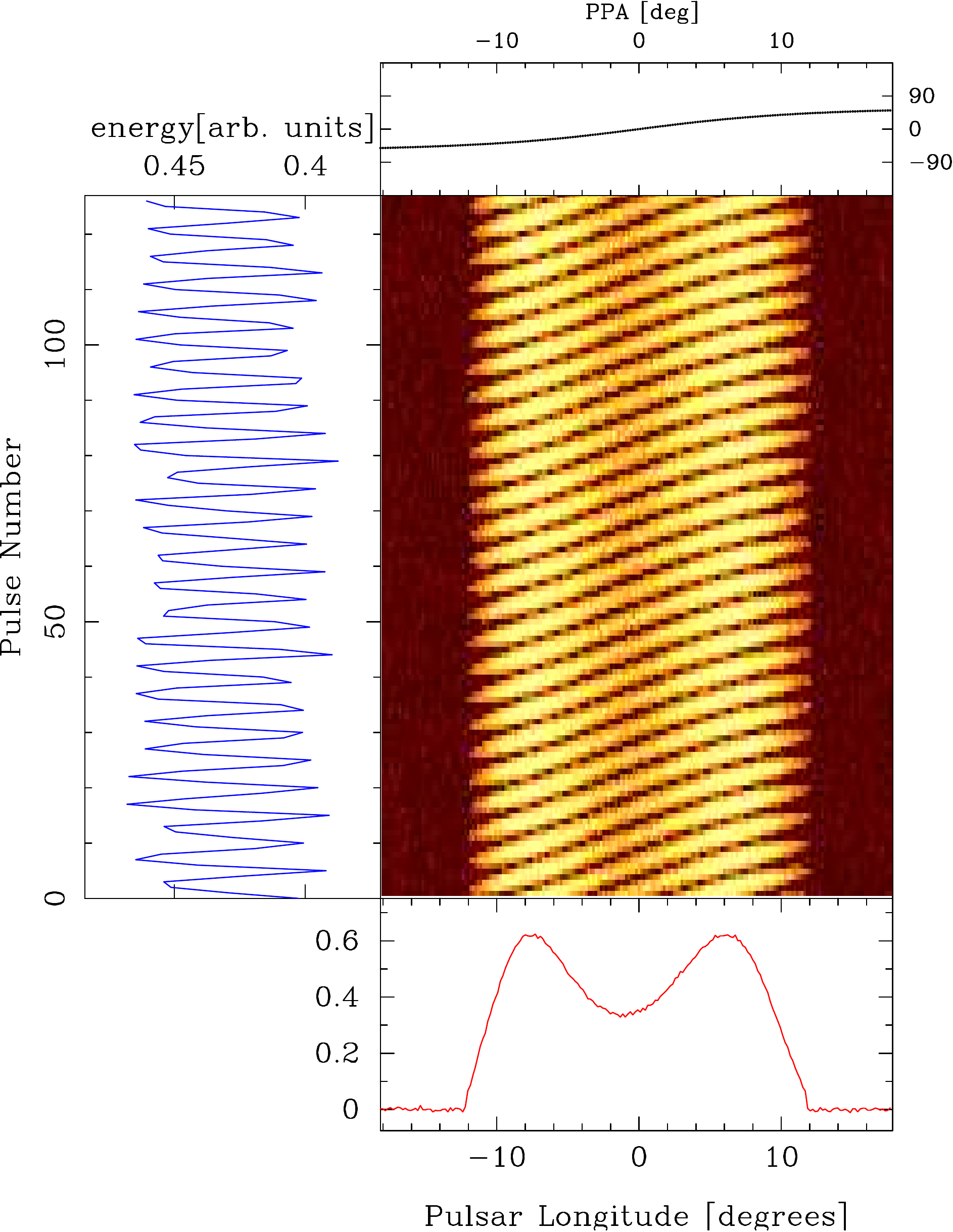}{0.48\textwidth}{(a)}
          \fig{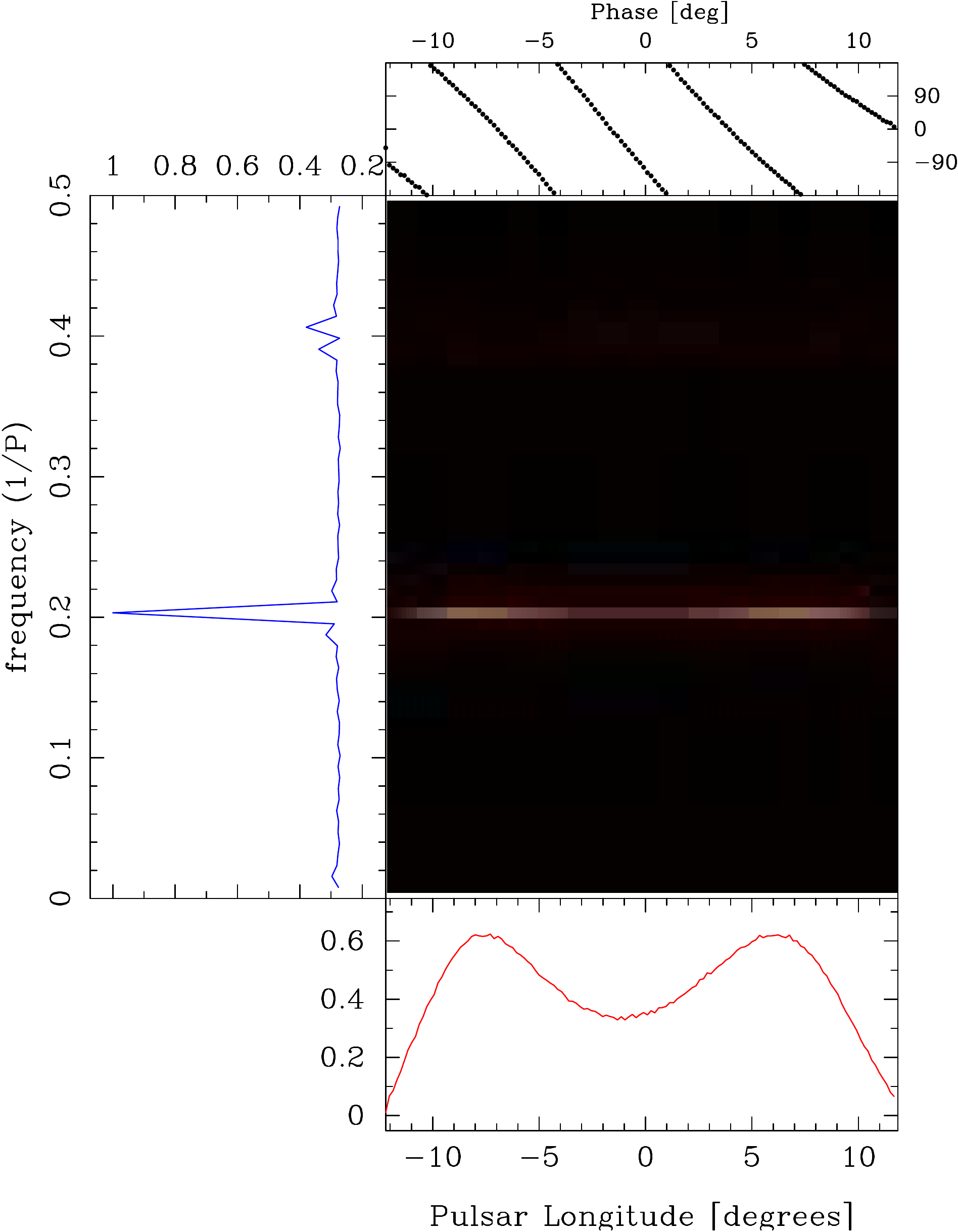}{0.48\textwidth}{(b)}}
\caption{The figure shows single pulse simulations demonstrating the coherent
subpulse drifting. (a) Pulse stack with 128 simulated single pulses and (b) 
Longitude Resolved Fluctuation Spectra (LRFS) across the pulse window. The 
drifting periodicity, $P_3$ = 4.9$P$ is seen as a peak frequency, $f_p\sim$0.2 
cycles/$P$. The time evolution of the sparking pattern is reflected in the 
phase behaviour across the profile (top window).
\label{fig:cohsingl}}
\end{figure}

We assume the emission to originate from an average height of $30R_S$ where the
opening angle of the open field line region is $\rho=4.55\degr$. In order to
simulate the single pulses exhibiting coherent drifting an outer LOS traverse 
is considered with $\beta=3\degr$, such that $\beta/\rho=0.66$. We have 
simulated 128 single pulses using the above setup and the corresponding pulse 
stack is shown in Fig. \ref{fig:cohsingl}a. The average profile shows a barely 
resolved double peaked structure resulting from the shifted non-dipolar polar 
cap and corresponding shift in the LOS cut across it (see Fig. 
\ref{fig:sparkcoh}). The single pulses show prominent drift bands with 
systematic variations across the entire window as expected for coherent 
drifting behaviour. The drifting behaviour is characterized using the Longitude
Resolved Fluctuation Spectra \citep[LRFS,][]{B73} as shown in Fig. 
\ref{fig:cohsingl}b. The LRFS is estimated by determining the FFT across the 
128 pulses along each pulsar longitude within the emission window. The drifting
periodicity is seen as the peak amplitude in frequency, $f_p=0.2$ cycles/$P$ 
(left window) while the evolution of the sparking pattern across the LOS is 
reflected in the large phase variations seen across the emission window (top 
window). The actual phase variations show continuous change from the leading to
the trailing edge of the profile, but has been wrapped around $\pm180\degr$ in 
the figure for convenience of plotting.

\subsection{Phase Stationary drifting}
\noindent
The phase-stationary drifting correspond to the cases where the subpulses do 
not show significant shift in position across the emission window but 
periodically changes in intensity. As a result the phases associated with the
drifting periodicity are relatively flat. The phase-stationary drifting is 
usually seen in pulsars with multiple components having core-cone profiles 
where the central core component do not exhibit any periodicity \citep{R86,
BMM19,BLK20}. Such profiles correspond to central LOS traverse of the emission 
beam with small $\beta$. We use the magnetic field configuration of Paper I to 
study this drifting behaviour where the the star centered dipole is specified 
as \textit{\textbf{d}} = ($d, 45\degr, 0\degr$) and there is one surface dipole
\textit{\textbf{m}} = ($0.05d, 0\degr, 0\degr$) located at \textit{\textbf{r}}
= ($0.95R_S, 57.08\degr, 20.66\degr$). The dipolar magnetic axis is tilted by
an angle 45$\degr$ with the rotation axis while the non-dipolar polar cap is
located around 20$\degr$ away from the dipolar polar cap. 

The polar cap is highly elliptical in shape and much smaller in size compared 
to the dipolar case, the fitting parameters shown in Table \ref{tab:capfit}. 
The surface magnetic field strength is characterized by average $b\sim19$ (see 
Fig. 18~in Paper I). Assuming $N_{trk} =2$, the typical spark size can be 
estimated as $a_{sprk}\sim13.9$ m, $b_{sprk}\sim4.0$ m and $h_{\perp}\sim7.5$ 
m. The screening factor for the PSG is obtained from eq.(\ref{eq:hperp}) as 
$\eta\sim0.021$. The average $\cos{\alpha_l}$ is 0.8 for the magnetic field 
configuration (Fig. 18~in Paper I) and we can estimate $P_3$ to be 9.7$P$.

\begin{figure}
\begin{interactive}{animation}{phase-stat_sprkcap.mp4}
\epsscale{0.55}
\plotone{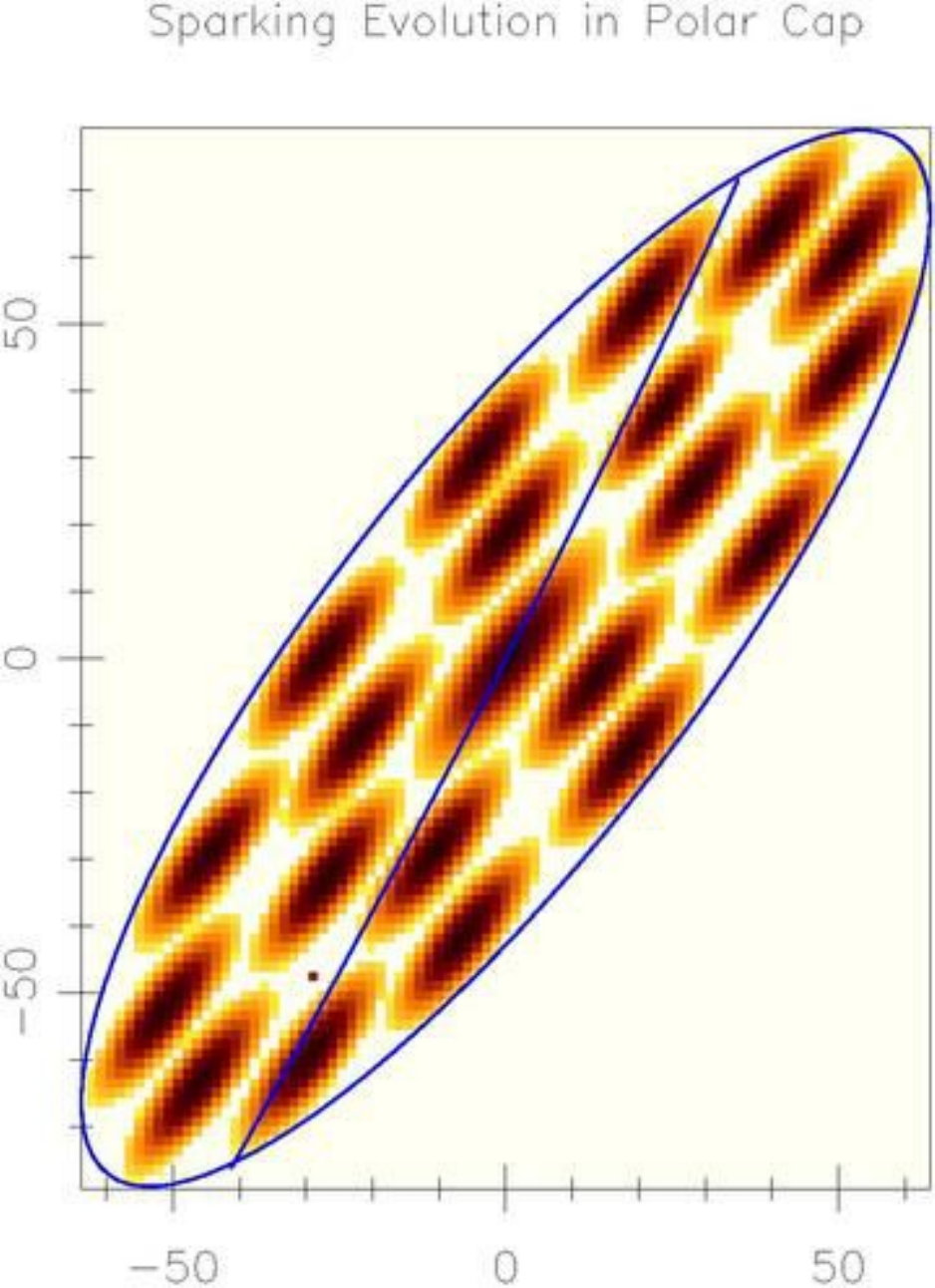}
\end{interactive}
\caption{The figure shows the two dimensional distribution of the sparking
pattern for an elliptical polar cap with major axis of 75.3 m, minor axis of
21.6 m and tilted by an angle of 52.4$\degr$ in the plane of the ellipse. The
sparks are arranged in two concentric elliptical annulus around a central spark
in a tightly packed configuration. The sparking pattern evolves with time to
show a clockwise shift in the left half and an anti-clockwise shift in the
right half bounded by the points $\theta'_s = 252.1\degr$ and $\theta'_e = 
72.1\degr$, where the pattern shifts away from $\theta'_s$ and converges
towards $\theta'_e$. The line of sight (LOS) traverse across the emission beam
centrally at an angle $\beta=-0.2\degr$ and its imprint on the polar cap is 
also shown. The dynamical evolution of the sparking distribution across the LOS
results in phase stationary subpulse drifting in the inner and outer concentric
rings, but no periodic behaviour in the central component. \\\\ An animation 
showing the evolution of the sparking configuration with time is available.
\label{fig:sparkphs}}
\end{figure}

\begin{deluxetable}{ccccccccccc}
\tablecaption{The details of the sparking distribution in polar cap exhibiting 
phase stationary drifting \label{tab:sparkphs}}
\tablewidth{0pt}
\tablehead{
    & \colhead{$i$} & \colhead{$a_{out}$} & \colhead{$a_{in}$} & \colhead{$b_{out}$} & \colhead{$b_{in}$} & \colhead{$N_{sprk}$} & \colhead{$\theta_{sprk}$} & \colhead{$a_{trk}$} & \colhead{$b_{trk}$} & \colhead{$\omega_{u,d}$}  \\
   &   & \colhead{(m)} & \colhead{(m)} & \colhead{(m)} & \colhead{(m)} &   & \colhead{($\degr$)} & \colhead{(m)} & \colhead{(m)} & \colhead{(deg s$^{-1}$)}
          }
\startdata
  Outer & 1 & 75.3 & 47.5 & 21.6 & 13.6 & 13 & 27.7 & 61.4 & 17.6 & $\mp$2.9 \\
  Inner & 2 & 47.5 & 19.7 & 13.6 & 5.6 & 7 & 51.4 & 33.6 & 9.6 & $\mp$5.3 \\
\enddata
\end{deluxetable}

The two dimensional distribution of the sparks and their evolution with time is
shown in Fig.\ref{fig:sparkphs}. The sparks are distributed around two 
elliptical rings surrounding a central spark in a tightly packed configuration.
The details of the sparking distribution for each track is reported in Table
\ref{tab:sparkphs} and shows the different major ($a$) and minor ($b$) axes
describing each ring, the maximum number of fully formed spark that can be
accommodated along these rings ($N_{sprk}$), the angular size of the sparks 
($\theta_{sprk}$), as explained in section \ref{sec:spark2D} and the rate of 
shifting of the patterns $\omega_{u,d}$. The two bounding points around which 
the sparking distribution show evolution in opposite directions are $\theta'_s 
= 252.1\degr$ and $\theta'_e = 72.1\degr$, where the pattern shifts away from 
$\theta'_s$ and converges towards $\theta'_e$. The average shift of the 
sparking pattern during a rotation period ($P=1$ s) is $h_D\sim1.7$ m. 

\begin{figure}
\gridline{\fig{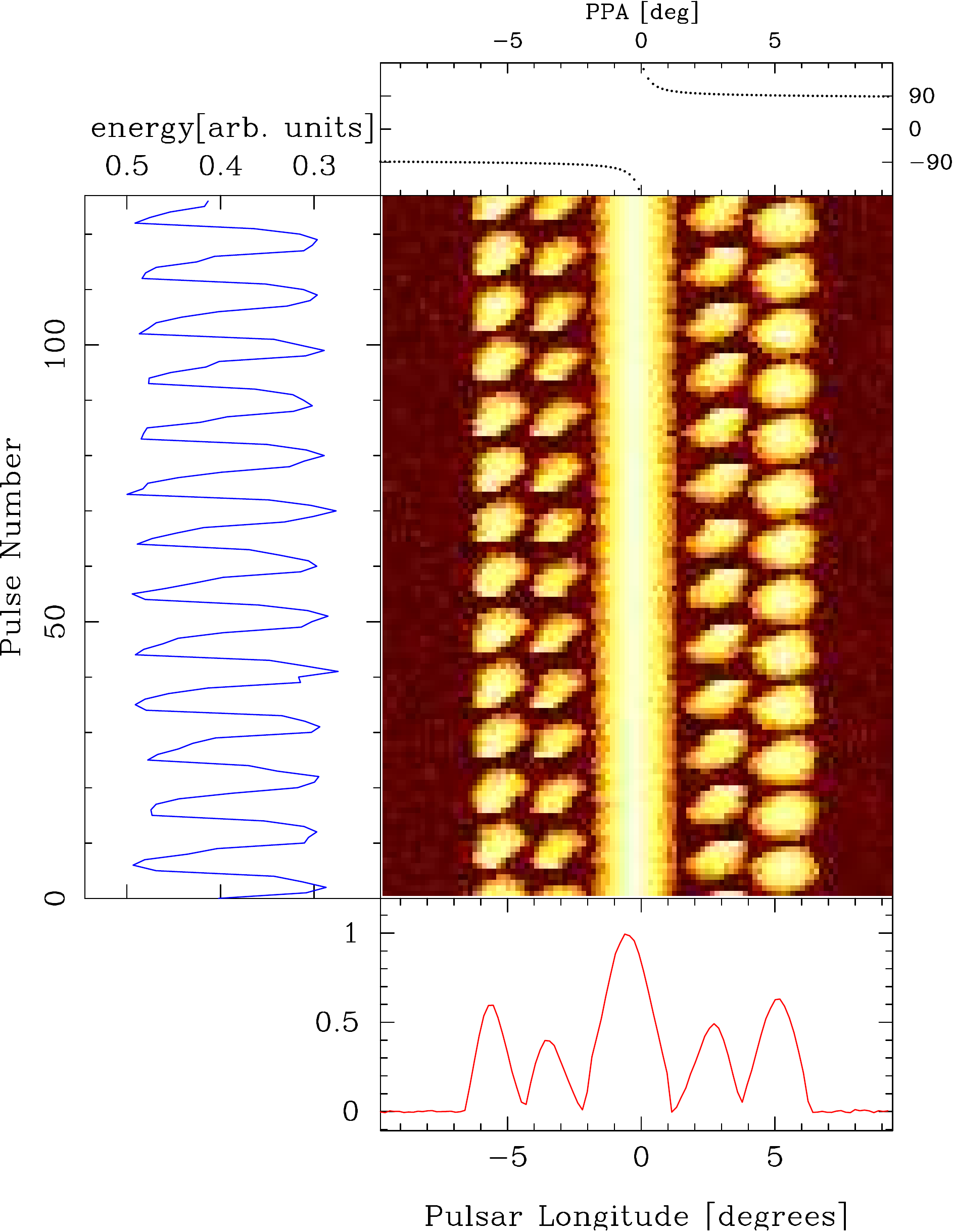}{0.48\textwidth}{(a)}
          \fig{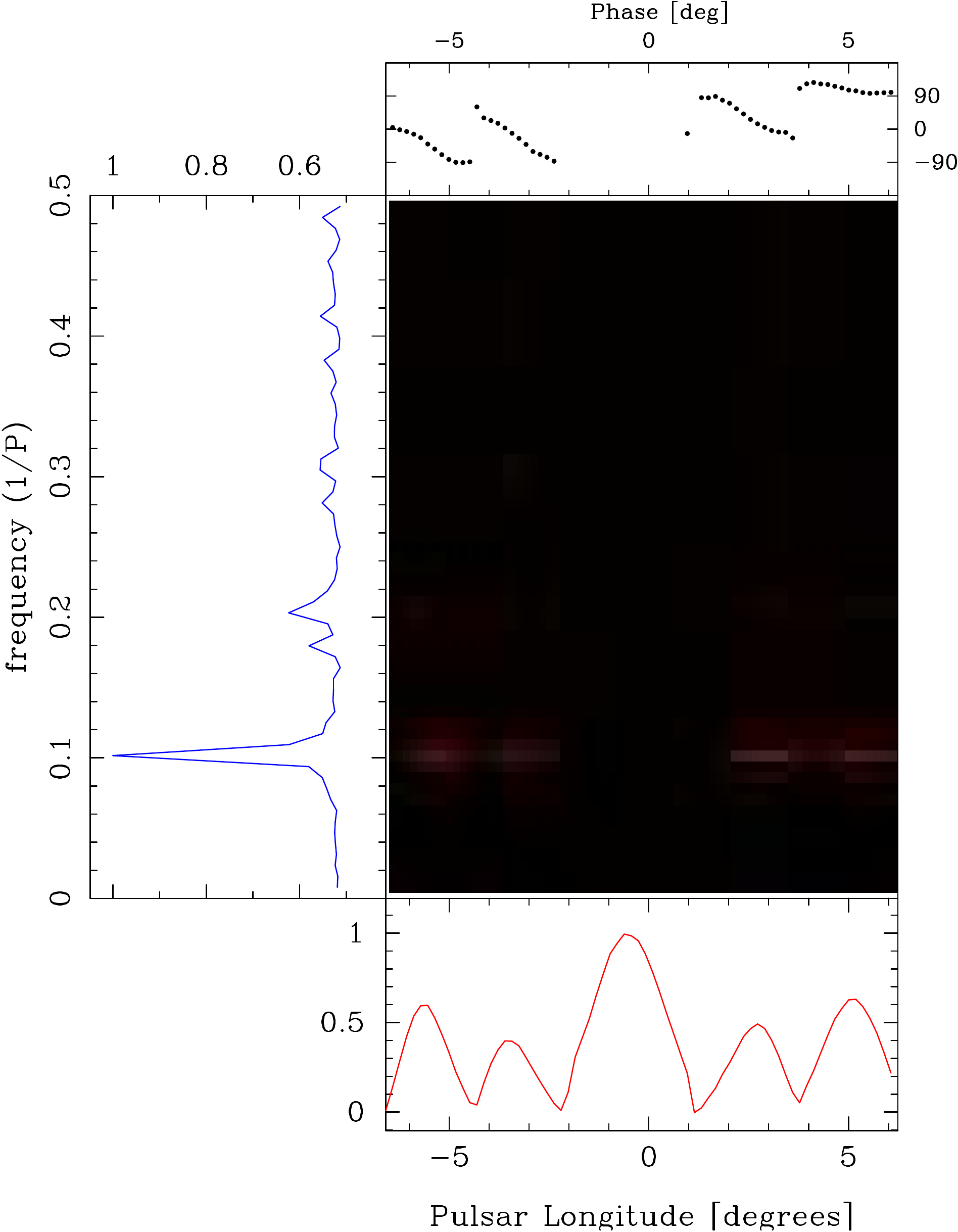}{0.48\textwidth}{(b)}}
\caption{The figure shows single pulse simulations demonstrating the phase 
stationary drifting. (a) Pulse stack with 128 simulated single pulses. There 
are five components in the average profile and the central core do not exhibit
any periodicity. (b) Longitude Resolved Fluctuation Spectra (LRFS) across the 
pulse window. The drifting periodicity, $P_3 = 9.7P$ is seen as a peak 
frequency, $f_p\sim0.1$ cycles/$P$. The time evolution of the sparking pattern
is reflected in the phase behaviour across the profile (top window).
\label{fig:phs_singl}}
\end{figure}

We assume the emission to originate from an average height of $30R_S$ where the
opening angle of the open field line region is $\rho=4.55\degr$. In order to
simulate the single pulses exhibiting phase-stationary drifting a central LOS 
traverse is considered with $\beta=-0.2\degr$, such that $\beta/\rho=0.04$. We 
have simulated 128 single pulses using the above setup and the corresponding 
pulse stack is shown in Fig. \ref{fig:phs_singl}a. The average profile has five
components where the central core component corresponds to the central region
with a constant presence of sparking. As a result the core component does not 
show any drifting which is an established observational result. The surrounding
components show phase stationary drifting with periodic change of intensity but
very little systematic variations across the longitudes. The drifting behaviour
is characterized using the LRFS as shown in Fig. \ref{fig:phs_singl}b. The 
drifting periodicity is seen as the peak amplitude in frequency, $f_p\sim0.1$ 
cycles/$P$ (left window) while small scale phase variations are seen across 
each component (top window). 

\subsection{Bi-drifting : Reversals in Drift directions}
\noindent
The Bi-drifting phenomenon is a unique drifting behaviour seen in a handful of
pulsars \citep{CLM05,W16,SvL17,BM18a,BPM19,SvLW20,SBD22} where the drifting
direction is opposite in different components of the pulsar profile. These 
pulsars usually have wider profiles that suggest a small inclination angle 
$\theta_d$ between the rotation and the magnetic axis. The magnetic field 
configuration used for simulating Bi-drifting in Paper I consisted of a star 
centered dipole specified as \textit{\textbf{d}} = ($d, 5\degr, 0\degr$) and 
one surface dipole \textit{\textbf{m}} = ($0.005d, 0\degr, 0\degr$) located at 
\textit{\textbf{r}} = ($0.95R_S, 5\degr, 120\degr$). The dipolar magnetic axis 
has a low inclination angle of 5$\degr$ with the rotation axis, while the 
non-dipolar polar cap is rotated by a large angle of around 100$\degr$ away 
from the dipolar polar cap.

The polar cap is elliptical in shape and is tilted in the opposite direction
compared to the previous cases. The parameters of the elliptical fit to the
polar cap outline are reported in Table \ref{tab:capfit}. The surface magnetic 
field strength is characterized by average $b\sim26$ (see Fig. 19~in Paper I). 
Assuming $N_{trk} =2$, the typical spark size can be estimated as 
$a_{sprk}\sim11.5$ m, $b_{sprk}\sim6.7$ m and $h_{\perp}\sim8.8$ m. The 
screening factor for the PSG is obtained from eq.(\ref{eq:hperp}) as 
$\eta\sim0.013$. The average $\cos{\alpha_l}$ is 0.79 for the magnetic field
configuration (Fig. 19~in Paper I) and the expected $P_3$ is 15.7$P$.

\begin{figure}
\begin{interactive}{animation}{bidrift_sprkcap.mp4}
\epsscale{0.55}
\plotone{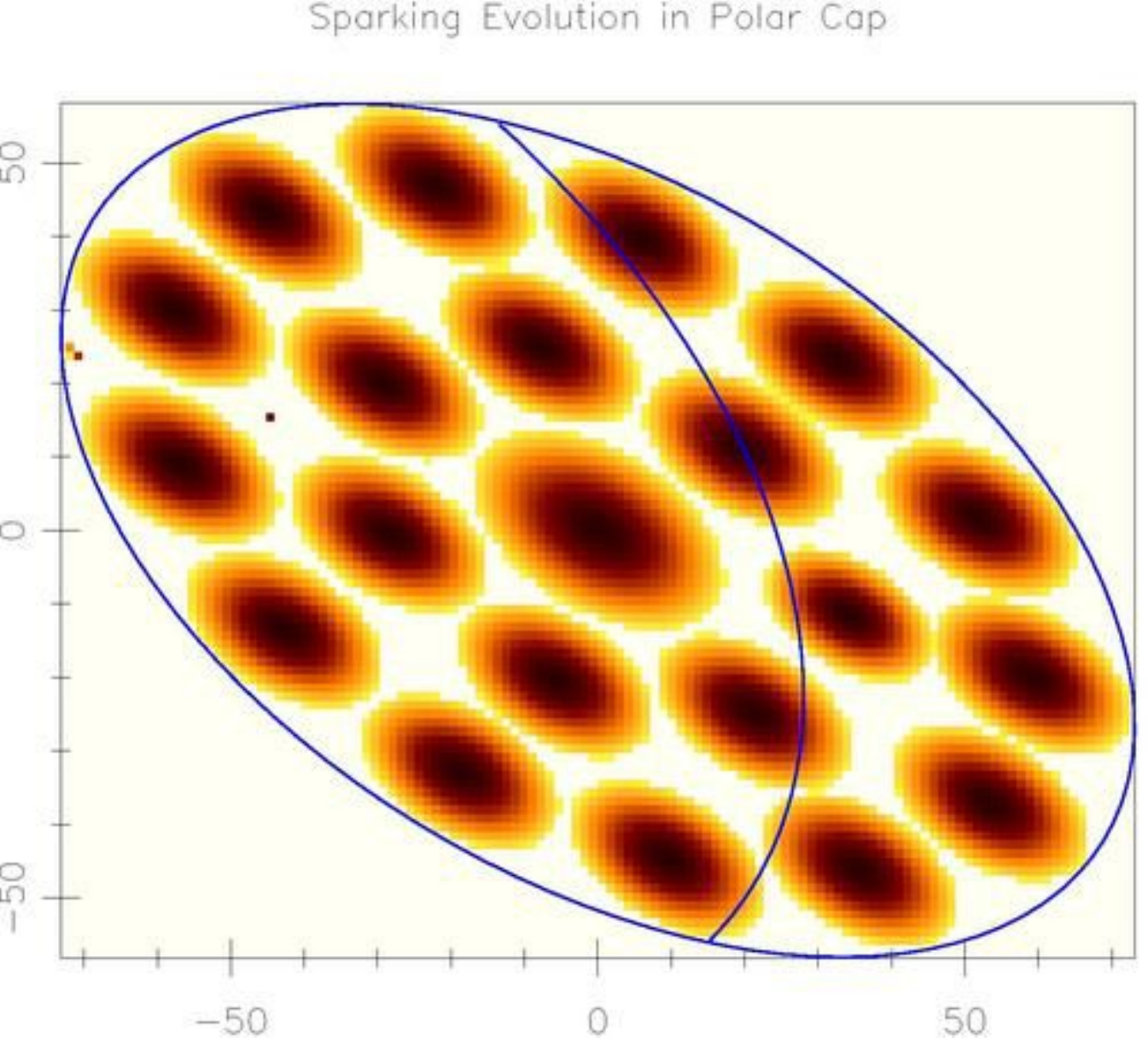}
\end{interactive}
\caption{The figure shows the two dimensional distribution of the sparking
pattern for an elliptical polar cap with major axis of 62.2 m, minor axis of
36.1 m and tilted by an angle of 148.5$\degr$ in the plane of the ellipse. The
sparks are arranged in two concentric elliptical annulus around a central spark
in a tightly packed configuration. The sparking pattern evolves with time to
show a clockwise shift in the upper half and an anti-clockwise shift in the
lower half bounded by the points $\theta'_s = 169.9\degr$ and $\theta'_e = 
-10.1\degr$, where the pattern shifts away from $\theta'_s$ and converges
towards $\theta'_e$. The line of sight (LOS) traverses across the emission beam
between the center and the edge at an angle $\beta=2\degr$ and its imprint on 
the polar cap is also shown. The dynamical evolution of the sparking 
distribution across the LOS results in Bi-drifting with opposite drift 
directions in the upper and lower halves. \\\\ An animation showing the 
evolution of the sparking configuration with time is available.
\label{fig:sparkbi}}
\end{figure}

\begin{deluxetable}{ccccccccccc}
\tablecaption{The details of the sparking distribution in polar cap exhibiting
Bi-drifting \label{tab:sparkbi}}
\tablewidth{0pt}
\tablehead{
    & \colhead{$i$} & \colhead{$a_{out}$} & \colhead{$a_{in}$} & \colhead{$b_{out}$} & \colhead{$b_{in}$} & \colhead{$N_{sprk}$} & \colhead{$\theta_{sprk}$} & \colhead{$a_{trk}$} & \colhead{$b_{trk}$} & \colhead{$\omega_{u,d}$}  \\
   &   & \colhead{(m)} & \colhead{(m)} & \colhead{(m)} & \colhead{(m)} &   & \colhead{($\degr$)} & \colhead{(m)} & \colhead{(m)} & \colhead{(deg s$^{-1}$)}
          }
\startdata
  Outer & 1 & 62.2 & 39.2 & 36.1 & 22.7 & 13 & 27.7 & 50.7 & 29.4 & $\mp$1.8 \\
  Inner & 2 & 39.2 & 16.2 & 22.7 & 9.3 & 7 & 51.4 & 27.7 & 16.0 & $\mp$3.3 \\
\enddata
\end{deluxetable}

The two dimensional distribution of the sparks and their evolution with time is
shown in Fig.\ref{fig:sparkbi}. The sparks are distributed around two 
elliptical rings surrounding a central spark in a tightly packed configuration.
The details of the sparking distribution for each track is reported in Table
\ref{tab:sparkbi} and shows the different major ($a$) and minor ($b$) axes
describing each ring, the maximum number of fully formed spark that can be
accommodated along these rings ($N_{sprk}$), the angular size of the sparks
($\theta_{sprk}$), as explained in section \ref{sec:spark2D} and the rate of
shifting of the patterns $\omega_{u,d}$. Due to the large shift in the location
of the non-dipolar polar cap the sparking distribution evolves along upper and 
lower parts of the concentric rings, in contrast with the previous cases where
the changes happened on the left and right halves. The two bounding points 
around which the sparking distribution show evolution in opposite directions 
are $\theta'_s = 169.9\degr$ and $\theta'_e = -10.1\degr$, where the pattern 
shifts away from $\theta'_s$ and converges towards $\theta'_e$. The average 
shift of the sparking pattern during a rotation period ($P=1$ s) is 
$h_D\sim1.2$ m.

\begin{figure}
\gridline{\fig{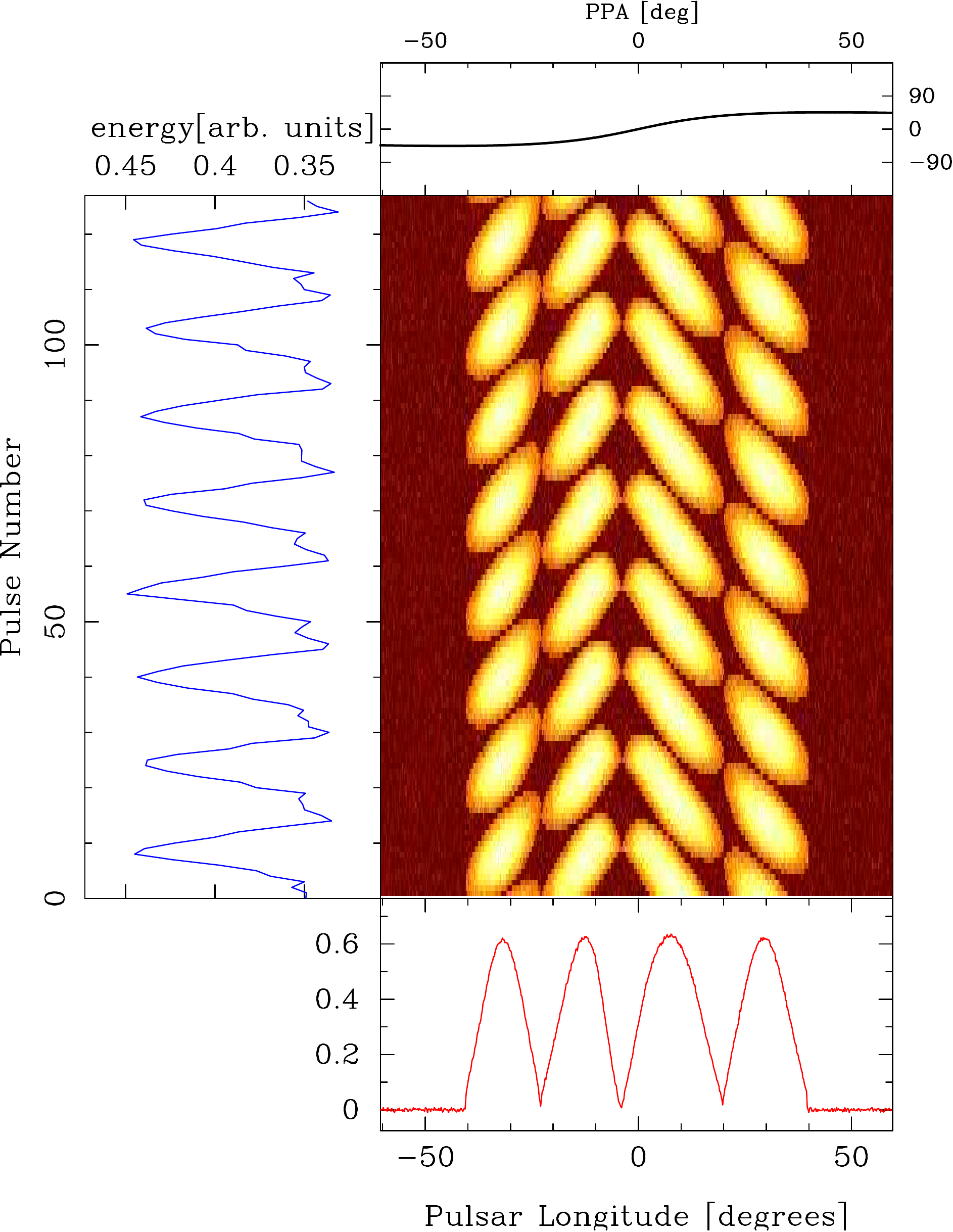}{0.48\textwidth}{(a)}
          \fig{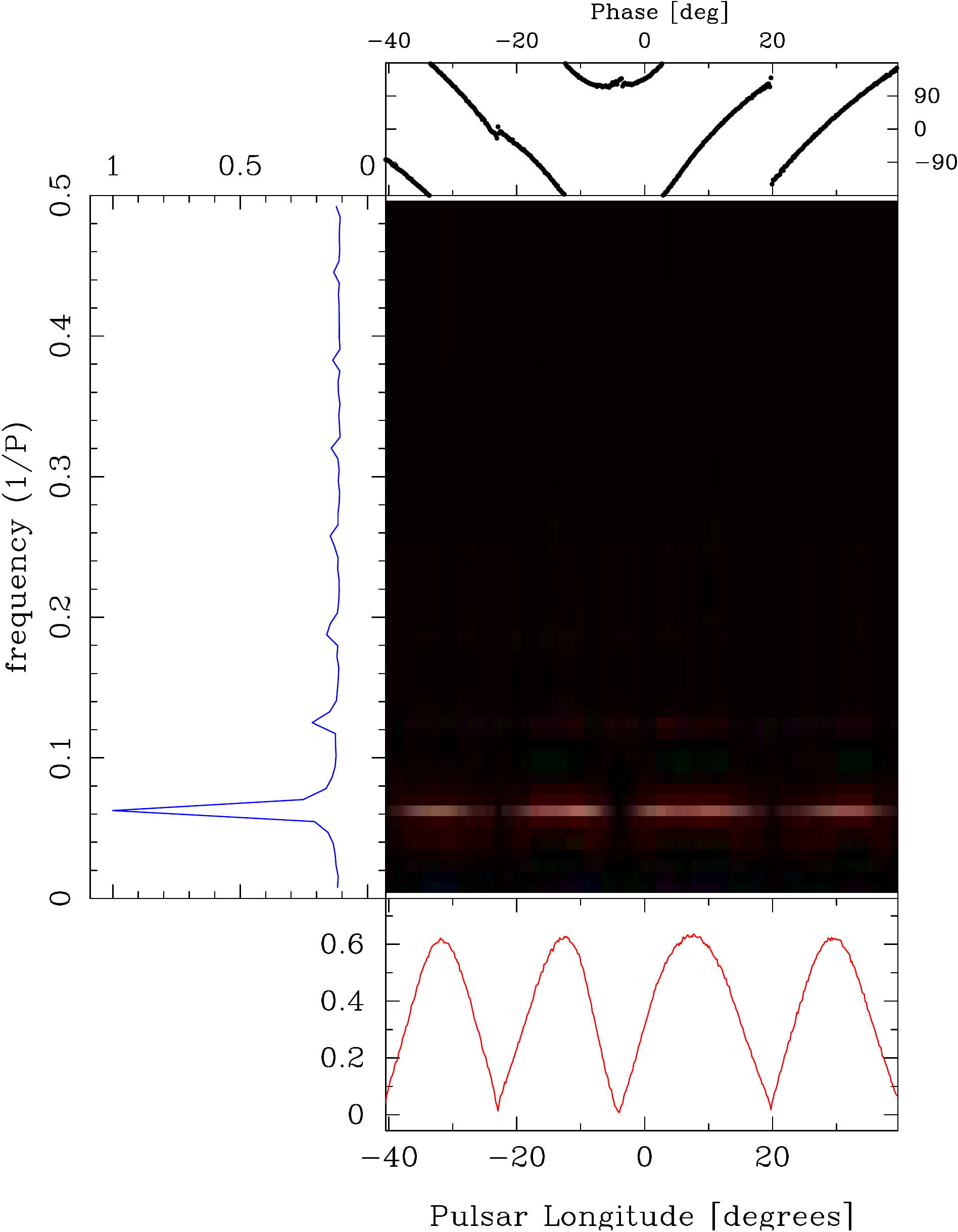}{0.48\textwidth}{(b)}}
\caption{The figure shows single pulse simulations demonstrating Bi-drifting 
behaviour. (a) Pulse stack with 128 simulated single pulses. There are four 
components in the average profile with the two leading components drifting in 
opposite direction to the trailing ones. (b) Longitude Resolved Fluctuation 
Spectra (LRFS) across the pulse window. The drifting periodicity, $P_3 = 15.7P$
is seen as a peak frequency, $f_p\sim0.06$ cycles/$P$, in the left window. The 
time evolution of the sparking pattern is reflected in the phase behaviour 
across the profile with the first two components showing a negative slope 
while the phases show a positive slope in their variations in the trailing 
components (top window).
\label{fig:bi_singl}}
\end{figure}

We assume the emission to originate from an average height of $30R_S$ where the
opening angle of the open field line region is $\rho=4.55\degr$. In order to
show the full effect of the Bi-drifting behaviour a LOS traverse halfway 
between the center and the edge is considered with $\beta=2\degr$, 
$\beta/\rho=0.4$, such that all conal components are visible without being 
affected by the stationary cone. In contrast to the previous two cases the LOS
is roughly perpendicular to the direction of the shifts in the sparking 
pattern. We have simulated 128 single pulses using the above setup and the 
corresponding pulse stack is shown in Fig. \ref{fig:bi_singl}a. The average 
profile has four components with prominent drifting behaviour where the leading
two components have opposite shifts in their drifting pattern compared to the 
trailing components. The drifting behaviour is characterized using the LRFS as 
shown in Fig. \ref{fig:bi_singl}b. The drifting periodicity is seen as the 
peak frequency, $f_p\sim0.06$ cycles/$P$ (left window), while phase changes 
show opposite slopes in the leading and trailing components (top window).

\section{Summary \& Conclusion} \label{sec:conclusion}
\noindent
We have presented a model for the two dimensional distribution of sparking 
discharges in the IAR above the pulsar polar cap and its evolution with time. 
The polar cap is dominated by non-dipolar magnetic fields and is smaller in
size and usually shifted compared to an equivalent dipolar polar cap. The 
surface of the polar cap is heated to temperatures $\sim10^6$ K, around the
critical temperature for ionic free flow, resulting in a partially screened 
gap. The sparking discharges in this system is setup when the temperature goes 
below the critical level opening up large potential difference for pair 
cascades to commence. The sparks are essentially a mechanism to regulate the 
surface temperature where the surface acts as a thermostat. The primary 
features of the sparking distribution are summarized as follows :
\begin{enumerate}[leftmargin=*]
\item The sparks are formed in a tightly packed configuration governed by the 
surface temperature. The sparks spread out around the point of origin depending
on the available potential drop across the field lines, with decreasing 
particle densities away from the peak location. The region lying between sparks
are also heated by diffusion of particles from surrounding sparks and there are 
no effective gaps between sparks.
\item The sparks have typical durations $\sim$10-100 $\mu$sec which is the time 
taken to reach the critical temperature. The sparking is a local event on the 
surface, governed by the thermostatic regulation, and is unaffected by distant
sparks. During the sparking process the charges within the spark lags behind
the co-rotation motion due to \textit{\textbf{E}}\textbf{x}\textit{\textbf{B}} 
drift. The surface cooling timescales $\sim$ 100 $n$sec, is much shorter than 
the spark duration which ensures that immediately after the cessation of 
sparking the next spark can be formed at a nearby point.
\item The evolution of the tightly packed sparking distribution in a thermally 
regulated surface depends on the boundary of the polar cap. No sparking 
discharges can take place in the closed field line region and hence thermal 
regulation requires a continuous presence of sparking around the boundary. As 
the sparks on the boundary drift opposite to the co-rotation direction during 
their lifetimes, the location of maximally heated points moves around the 
boundary leading to time evolution of the sparking pattern. The individual 
sparks are short lived and do not participate in any long term periodic 
behaviour. The locus of the heating points on the surface around which the 
subsequent sparking take place show a gradual but continuous shift with time.
\item The evolution of the sparking pattern on the boundary ensures that 
similar changes take place in the interior as well, since these sparks also 
affect the heating on the other side. The sparks are setup around concentric 
rings where two distinct direction of evolution arises bounded by points where 
the co-rotation direction is normal to the curvature of the boundary. In one 
half the pattern represents a clockwise shift while in the other half an 
anti-clockwise shift is setup. As the pattern shifts, gaps open up in the two 
normal points in each ring where smaller sparks arise to aid the thermal 
regulation. The evolution of the sparking pattern described above requires the 
presence of a polar cap boundary with regular curvature like an ellipse or a 
circle. In case of irregular shapes it is possible that only part of the polar 
cap shows evolution or the sparking pattern is stationary in case of extremely 
irregular shapes.
\item Due to differential shift of sparking patterns in two halves the heating 
location at the center remains stationary with sparks forming at the same 
place at regular intervals. This resembles the core component in the pulsar
profile which do not exhibit any drifting behaviour. 
\end{enumerate}
The evolution of the sparking process in IAR was used to simulate the subpulse 
drifting behaviour in pulsars. We considered three different surface magnetic
field configurations, outlined in Paper I, to reproduce the different classes 
of drifting observed in the pulsar population. In each case detailed 
characterisation of the sparking properties on the surface was carried out. 
These provide the template for constraining the physical characteristics of 
the IAR in pulsars where subpulse drifting has been measured and will be 
explored in future works. A complete determination of the IAR properties will
likely require additional information from X-ray observations to constrain the 
polar cap temperature and size.

\section*{Acknowledgments}
DM acknowledges the support of the Department of Atomic Energy, Government of 
India, under project no. 12-R\&D-TFR-5.02-0700. DM acknowledges funding from 
the grant ``Indo-French Centre for the Promotion of Advanced Research - 
CEFIPRA" grant IFC/F5904-B/2018. This work was supported by the grant 
2020/37/B/ST9/02215 of the National Science Centre, Poland.

\bibliography{reflist}{}
\bibliographystyle{aasjournal}

\end{document}